\documentclass[a4paper,11pt,showkeys]{article}
\pdfoutput=1 
\usepackage{jheppub} 
\usepackage{tikz}
\usepackage{slashed}
\usepackage{bbold}
\usepackage{ulem}
\usepackage{subfigure}
\usepackage{graphicx}
\usepackage{comment}
\usepackage[all]{xy}
\usepackage{multirow}
\usepackage{float}
\usepackage{mathtools}
\usepackage[export]{adjustbox}

\usepackage{color}

\newcommand \be{\begin{eqnarray}}
\newcommand \ee{\end{eqnarray}}

\numberwithin{equation}{section}

\DeclareMathOperator{\llangle}{\big\langle\hspace{-1.2mm}\big\langle\hspace{-.5mm}}
\DeclareMathOperator{\rrangle}{\hspace{-.5mm}\big\rangle\hspace{-1.2mm}\big\rangle}
\DeclareMathOperator{\Tr}{Tr}

\DeclareMathOperator{\arctanh}{arctanh}

\DeclareMathOperator{\diag}{diag}

\newcommand{\bea}{\begin{eqnarray}}
\newcommand{\eea}{\end{eqnarray}}
\newcommand{\beq}{\begin{equation}}
\newcommand{\eeq}{\end{equation}}
\newcommand{\bal}{\begin{equation}\begin{aligned}}
\newcommand{\eal}{\end{aligned} \end{equation}}

\newcommand{\cA}{{\mathcal A}}

\newcommand{\cC}{{\mathcal C}}
\newcommand{\cD}{{\mathcal D}}

\newcommand{\cL}{{\mathcal L}}

\newcommand{\cN}{{\mathcal N}}
\newcommand{\cP}{{\mathcal P}}

\newcommand{\cO}{{\mathcal O}}

\newcommand{\cW}{{\mathcal W}}

\usepackage{hyperref}

\title{Wilson loops and defect RG flows in ABJM}

\author[a]{Luigi Castiglioni,}
\author[a]{Silvia Penati,}
\author[a]{Marcia Tenser,}
\author[b,c,d]{Diego Trancanelli}

\affiliation[a]{Dipartimento di Fisica, Universit\`a degli Studi di Milano--Bicocca and INFN, Sezione di Milano--Bicocca, Piazza della Scienza 3, 20126 Milano, Italy}
\affiliation[b]{Dipartimento di Scienze Fisiche, Informatiche e Matematiche, Universit\`a di Modena e Reggio Emilia, via G. Campi 213/A, 41125 Modena, Italy}
\affiliation[c]{INFN Sezione di Bologna, via Irnerio 46, 40126 Bologna, Italy}
\affiliation[d]{Institute of Physics, University of S\~ao Paulo, 05314-970 S\~ao Paulo, Brazil}

\emailAdd{l.castiglioni8@campus.unimib.it}
\emailAdd{silvia.penati@mib.infn.it}
\emailAdd{marciatenser@gmail.com}
\emailAdd{dtrancan@gmail.com} 

\abstract{
We continue our study of renormalization group (RG) flows on Wilson loop defects in ABJM theory, which we have initiated in {\tt arXiv:2211.16501}. We generalize that analysis by including non-supersymmetric fixed points and RG trajectories. To this end, we first determine the ``ordinary", non-supersymmetric Wilson loops, which turn out to be two and to include an R-symmetry preserving coupling to the scalar fields of the theory, contrary to their four-dimensional counterpart defined solely in terms of the gauge field holonomy. We then deform these operators by turning on bosonic and/or fermionic couplings, which trigger an elaborate, multi-dimensional network of possible RG trajectories connecting a large spectrum of fixed points classified in terms of the amount (possibly zero) of supersymmetry and R-symmetry preserved. The $\beta$-functions are computed to leading order in the ABJM coupling but exactly in the deformation parameters, using an auxiliary one-dimensional theory on the defect and a dimensional regularization scheme.
A striking result is the different behavior of the two ordinary Wilson loops, of which one turns out to be a UV unstable point while the other is IR stable. The same is true for the two 1/2 BPS Wilson loops. We interpret our results from a defect CFT (dCFT) point of view, computing the anomalous dimensions of the operators associated to the deformations and establishing appropriate g-theorems. In particular, the fermionic unstable fixed point is associated to a dCFT which is not reflection positive.
}

\arxivnumber{}

\begin{document}
\maketitle

\section{Introduction}

Three-dimensional supersymmetric Chern-Simons-matter theories, like ABJ(M) \cite{Aharony:2008ug,Aharony:2008gk} and other $\cN\ge 2$ quiver theories \cite{Gaiotto:2008sd,Imamura:2008dt,Hosomichi:2008jd,Hama:2010av}, have a rich moduli space of BPS Wilson loops  discovered along the years in, for example, \cite{Drukker:2008zx,Chen:2008bp,Kluson:2008zrv,Rey:2008bh,Drukker:2009hy,Gaiotto:2007qi,Ouyang_2015,Cooke_2015,Ouyang:2015iza,Ouyang:2015bmy,Mauri:2017whf,Mauri_2018,drukker2020bps,Drukker:2020dvr,Drukker:2022ywj,Drukker:2022bff,Castiglioni:2022yes}.\footnote{See \cite{Drukker:2019bev} for a fairly recent review.} Such Wilson loops typically come in families related by a certain number of parameters, that allow to interpolate continuously among representatives preserving varying amounts of supercharges of the theory. 

These interpolations can be studied from the point of view of RG flows on defects, as initiated recently in \cite{Castiglioni:2022yes} for ABJM, building on similar analyses done in four dimensions in \cite{Polchinski_2011} and subsequent literature, see for example \cite{Beccaria:2017rbe,Beccaria:2018ocq,Correa:2019rdk,Cuomo:2021rkm,Beccaria:2021rmj,Beccaria:2022bcr,Garay:2022szq,Aharony:2022ntz}. The main idea is to start from a certain loop operator -- a UV fixed point -- and deform it with a marginally relevant parameter, that either leads to another loop operator -- an IR fixed point -- or to infinity along runaway directions.

The most studied and best understood example is the interpolation, initially proposed in 
\cite{Polchinski_2011}, between the non-BPS Wilson loop of four-dimensional ${\cal N}=4$ super Yang-Mills (SYM) and the 1/2 BPS operator of that theory. This interpolation is controlled by a single parameter $\zeta$ that acts as a marginally relevant deformation of the non-BPS operator defined only in terms of the gauge field (for $\zeta=0$) and triggers a flow toward the supersymmetric operator \cite{Maldacena_1998} which is also coupled to a scalar of the theory (for $\zeta=1$).

Performing a similar investigation in the ABJM theory, one expects to find a much richer spectrum of flows, given the larger number of Wilson loops that can be defined. 
In \cite{Castiglioni:2022yes} our focus has been on interpolations and RG trajectories associated to operators in ABJM that always preserve some amount of supersymmetry, at least one supercharge, something we dubbed as `enriched flows'. Here we continue that analysis by including also fixed points and RG trajectories corresponding to non-supersymmetric operators, in the original spirit of \cite{Polchinski_2011}. 

The first step consists in identifying what are such non-BPS operators in the case of ABJM. Differently from the four-dimensional counterpart, these operators are not just defined in terms of the gauge fields, but they also include an R-symmetry preserving coupling to the scalars. This can be justified as follows. First of all, scalar bilinears have classical dimension 1 in three dimensions and allow for the contraction of the R-symmetry indices to produce the $SU(4)$ singlet operator $C_I \bar C^I$ (with $I=1,\ldots, 4$). It follows that when we perform the renormalization of the gauge field on the defect theory, nothing prevents it from mixing with the singlet. In fact, one discovers that, due to the interaction with the bulk theory, this new effective vertex is produced. It is divergent and can only be cancelled if the defect theory defined by the Wilson loop includes a coupling to the scalars as the one above. It turns out there are two such operators, which we call $W^\pm$, corresponding to the two possible signs in front of $C_I \bar C^I$. We demonstrate this in section~\ref{sec:ordinary}.\footnote{Similar results were found in \cite{Gabai:2022vri,Gabai:2022mya}, where a classification of line operators in Chern-Simons theories with matter was obtained.} 

Having identified the non-BPS, or ``ordinary'', Wilson loops we then proceed to deforming them in several ways and in computing the associated $\beta$-functions at leading order in the ABJM coupling and in the planar limit. We start in section~\ref{sec:bosonic} with deforming the scalar couplings only.
We restrict these deformations to be diagonal, thus generally given by four independent parameters that allow to either break (completely or in part) the R-symmetry group or to preserve it. We obtain the $\beta$-functions associated to these deformations using the same one-dimensional effective approach originally developed in \cite{Samuel:1978iy,Gervais:1979fv} for QCD and then extended to ABJM in \cite{Castiglioni:2022yes}.

As is well-known, BPS Wilson loops in three-dimensional quiver theories are defined in terms of superconnections including couplings to the fermions of the theory. It is then natural to also consider fermionic deformations, which we do in section \ref{sec:fer1} for purely fermionic deformations and in  section \ref{sec:generic} for more generic deformations of both fermions and scalars.

This whole setup allows for a very rich, multi-dimensional space of possible RG flows, which we depict in a series of plots. In these plots we denote fixed points according to their R-symmetry structure: $SU(4)$ invariants as squares, $SU(3)$ as triangles and $SU(2)$ as circles. When they are black no supersymmetry is preserved, whereas when they are colored they preserve some amount of supersymmetry: blue points are $1/6$ BPS and red points are $1/2$ BPS. There are multiple facets in these series of plots and we leave a detailed presentation for the body of the text.

\begin{figure}[ht]
    \centering \includegraphics[width=.4\textwidth]{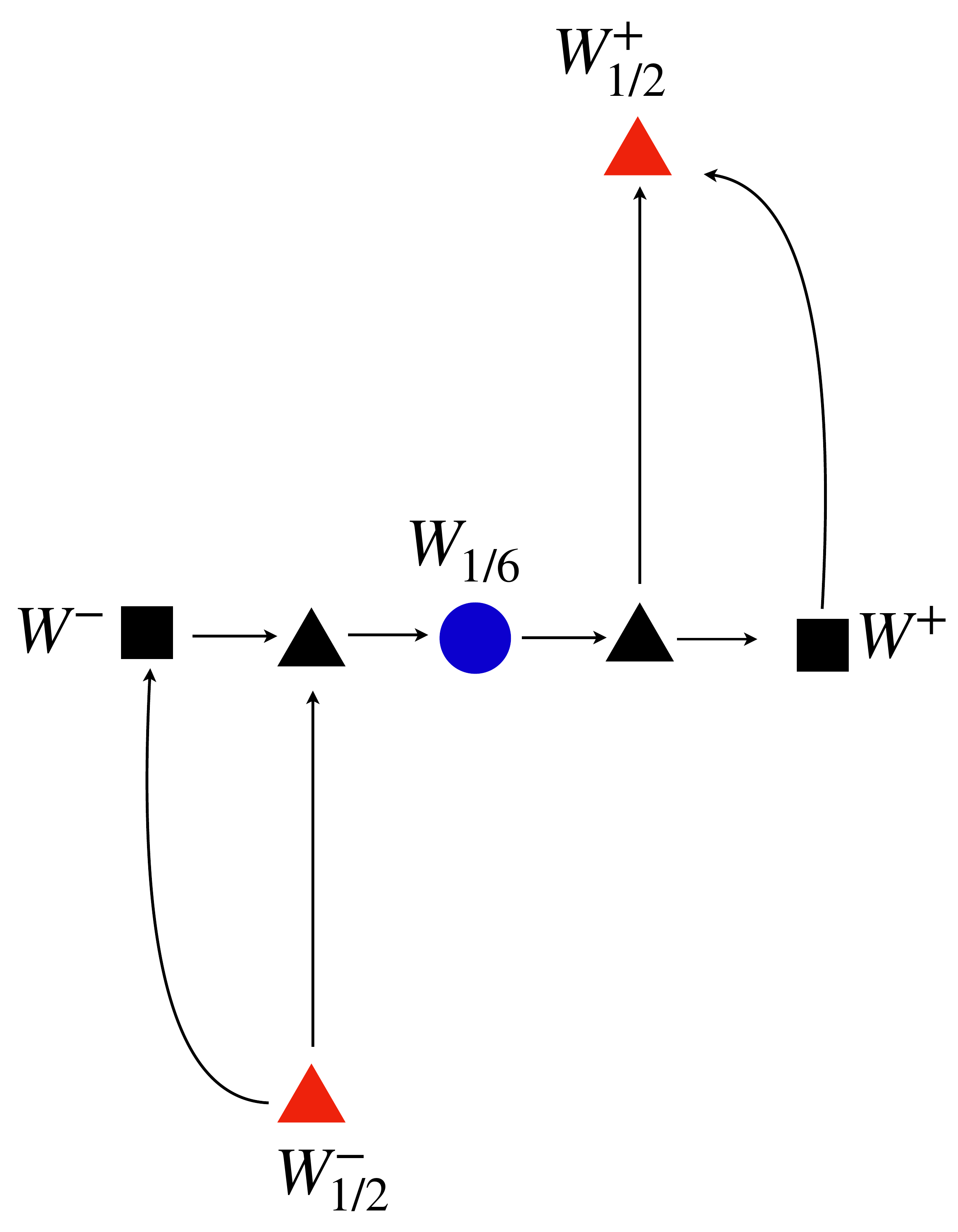}
    \caption{Representation of flows connecting different Wilson loop operators in ABJM theory. Along the horizontal line we have purely bosonic deformations whereas in the vertical direction we turn on fermions. Mixed deformations correspond to compositions of these two.
    }
\label{fig:schematics}
\end{figure}

Nevertheless, in figure \ref{fig:schematics} we show a glimpse of a schematic representation of some of our results. Along what we depict as a horizontal line we find bosonic flows. In particular, connecting the two $SU(4)$ symmetric ``ordinary" operators $W^-$ and $W^+$  at the end points (\begin{tikzpicture}
\fill[black] (0,0) rectangle (0.23,0.23);
  \end{tikzpicture}) we have two $SU(3)$ non-BPS bosonic loops (\begin{tikzpicture}
  \draw[black,fill=black] (0,0) -- 
  ++(0.25,0) --
  ++(-0.125,0.25) -- cycle;
  \end{tikzpicture}) and/or a bosonic $1/6$ BPS operators (\begin{tikzpicture}
  \filldraw [blue] (0,0) circle (3pt);
  \end{tikzpicture}), which is $SU(2)\times SU(2)$ symmetric. Along the vertical direction we represent fermionic flows and among these we find in particular two $1/2$ BPS fixed points (\begin{tikzpicture}
  \draw[red,fill=red] (0,0) -- 
  ++(0.25,0) --
  ++(-0.125,0.25) -- cycle;
  \end{tikzpicture}).
In this representation, arrows go from the UV to the IR. We hope the reader carries this general picture in mind throughout the paper.
  
One of the most interesting outputs of this web of flows is the qualitatively different behavior between $W^+$ and $W^-$, and between $W_{1/2}^+$ and $W_{1/2}^-$. Though classically they are equivalent (they simply differ by harmless signs in the couplings, which do not affect symmetries), at quantum level they describe completely different defect theories. In fact, while $W^-$ is a UV unstable fixed point, its counterpart $W^+$ is IR stable. Similarly, including fermionic deformations we find that UV instability lies at $W_{1/2}^-$ while IR stability corresponds to $W_{1/2}^+$.

After unravelling this multi-dimensional space of possible flows, in section \ref{sec:defecttheory} we interpret our results from the point of view of a defect CFT. First of all, computing anomalous dimensions at the first non-trivial order, we find that the perturbations are marginally relevant operators for the  fermionic $W^-$ and $W^-_{1/2}$ defects, consistently with the direction of the flows. 
Then, through the evaluation of 
the defect entropy \cite{Affleck:1991tk}, we manage to test the validity of a g-theorem \cite{Affleck:1991tk,Friedan:2003yc,Casini:2016fgb,Cuomo:2021rkm,Casini:2022bsu}. 

A striking result arises, which concerns the g-theorem along the flows connecting $W^\pm_{1/2}$ defect
theories with non-BPS bosonic $SU(3)$ invariant defects. We find that, while $W^+_{1/2}$ satisfies reflection-positivity and the g-theorem \cite{Cuomo:2021rkm} holds, the $W^-_{1/2}$ defect does not because of  a crucial sign change in the two-point function of its stress-tensor. As a consequence, the RG flow is still irreversible, but the defect entropy is increasing rather than decreasing from the UV unstable to the IR stable fixed points. However, this seems to be consistent with the dual description of defects at strong coupling and the Higgsing procedure used to construct these two operators \cite{Lietti:2017gtc}. In fact, in field theory they arise by Higgsing either with particle or antiparticle modes, while in M-theory $W^+_{1/2}$ is dual to a M2-brane configuration and $W^-_{1/2}$ is dual to an anti-M2 brane. This might explain why one defect is reflection-positive, whereas the other one is not.

We conclude with section \ref{sec:conclusions},
where we further discuss our findings and address a few relevant open questions. The technical details of our calculations are reported in four appendices. 
 
As a final remark, we stress that the same caveats of \cite{Castiglioni:2022yes} regarding the employed regularization scheme also apply here. All our computations are in fact performed in dimensional regularization, which is alternative to introducing framing and can then be thought as giving results which correspond to framing zero. Restricting to BPS Wilson loops, this is of course not what one obtains from localization, which works at framing one. However, it is well-known that the results from the two different schemes differ simply by an overall phase, the famous framing factor \cite{Witten:1988hf, Guadagnini:1989am, Alvarez:1991sx} or framing function \cite{Bianchi:2016yzj}. Since dimensional regularization breaks conformal invariance, while localization requires and preserves superconformal invariance \cite{Pestun_2012,Kapustin:2009kz}, the framing function has the interpretation of a conformal (or framing) anomaly. It is ultimately responsible for the BPS RG flows that we find at framing zero, thus we conclude that enriched flows are anomaly driven. In the more general case of non-BPS flows, the absence of cohomological equivalence does not guarantee that this is the case.
\section{Ordinary Wilson loops in ABJM}
\label{sec:ordinary}

Following what has been done in $\cN=4$ SYM theory, see for instance \cite{Polchinski_2011,Beccaria:2017rbe,Beccaria:2018ocq,Beccaria:2019dws,Beccaria:2021rmj,Beccaria:2022bcr}, here we want to study how to generalize to ABJM theory the idea of interpolating between the non-supersymmetric Wilson loop (WL) given by the holonomy of the gauge field (usually referred to as ``ordinary") and BPS loops that include extra couplings to the matter fields of the theory\footnote{A brief summary of the ABJM Wilson loops entering our study is given in appendix \ref{app:BPSWL}. We will always refer to this appendix for details on their explicit realization.}.

In the $\cN=4$ SYM case, the interpolation takes the form \cite{Polchinski_2011}
\beq
\label{eq:N=4WL}
W^{(\zeta)}(C)=\frac{1}{N} \Tr \cP \exp \oint_C d\tau\left[ \, i{A}_\mu(x) \dot{x}^\mu + \zeta \Phi_m(x)\theta^m\vert\dot{x}\vert \,\right], \qquad \theta^m \theta^m =1, 
\eeq
where $m$ runs over the six scalar fields of the theory. 

This is a one-parameter family of Wilson loops that connects the ordinary WL at $\zeta=0$ and the 1/2 BPS Wilson-Maldacena loop \cite{Maldacena_1998} at $\zeta=1$. Since $\langle W^{(\zeta)}(C) \rangle$ is invariant under $\zeta \to -\zeta$, one can restrict to $\zeta \geq 0$.

In \cite{Polchinski_2011} it was found perturbatively at first order in the 't Hooft coupling $\lambda$ that a non-trivial RG flow connecting these two operators exists, which is dictated by the following $\beta$-function
\beq
\beta_\zeta = \mu \frac{d\zeta}{d\mu} = -\frac{\lambda}{8\pi^2} \zeta(1-\zeta^2) + \cO(\lambda^2).
\eeq
Around the $\zeta=0$ UV fixed point the scalar perturbation $\Phi_m \theta^m$ becomes marginally relevant and drives the system towards the 1/2 BPS IR fixed point at $\zeta =1$. 

We note that within the one-parameter family \eqref{eq:N=4WL} the ordinary Wilson loop corresponds to the operator preserving the maximal amount of R-symmetry, that is $SO(6)$. The $\zeta$-deformation breaks $SO(6)$ to $SO(5)$. The Polchinski-Sully flow can then be interpreted as connecting the maximally R-symmetric operator to the maximally supersymmetric one. 

In order to investigate  the existence of an analogous pattern in ABJM theory, we focus on the study of parametric deformations that connect non-BPS with BPS Wilson loops.
The logic that we are going to apply is the following: We start with the maximally R-symmetric operator, that is the ``ordinary" WL preserving the maximal amount of R-symmetry, we then add marginally relevant, partial R-symmetry breaking  deformations and study the corresponding RG flows.  

Naively, one would expect the ``ordinary" WL in the ABJM theory to be the same as for ${\cal N}=4$ SYM, that is the gauge field holonomy given by  \eqref{eq:N=4WL} with $\zeta=0$. In fact, $A_{\mu} \dot x^{\mu}$ is invariant under the action of the full R-symmetry group, $SO(6) \simeq  SU(4)$. However, this is not correct, since 
in this case there are other R-symmetry preserving operators with the same dimension, which necessarily mix with $A_{\mu} \dot x^{\mu}$ and need to be included.

In order to prove that, we make use of the auxiliary one-dimensional method originally proposed for QCD in \cite{Dorn:1986dt,Samuel:1978iy,Gervais:1979fv,Arefeva:1980zd,CRAIGIE1981204} and suitably generalized to the ABJM case in \cite{Castiglioni:2022yes}.
As we are going to show, in this approach the appearance of further R-symmetry preserving operators is signaled by the fact that the $A_{\mu} \dot x^{\mu}$ operator itself leads to a non-renormalizable one-dimensional auxiliary theory.

As done in \cite{Castiglioni:2022yes} (see also \cite{Dorn:1986dt, CRAIGIE1981204}),  we take the loop to be supported along the circular curve parametrized as
\beq
\label{eq:xmu}
x^\mu(\tau) = (0,\cos\tau,\sin\tau)\,, \quad \tau \in [0,2\pi)\,.
\eeq
For the ordinary Wilson loop we introduce the one-dimensional fermionic field $z$($\bar z$) in the (an\-ti)\-fun\-da\-men\-tal representation of $U(N)$ and define the Wilson loop vacuum expectation value (VEV) on a contour $\cC_{12}$ connecting two points $\tau_1,\tau_2$ as\footnote{This prescription is briefly reviewed in appendix \ref{app:ren}. It is easy to realize that for the ordinary Wilson loop the $\Psi$ auxiliary fermion defined in \eqref{eq:oddmatrix} boils down to $z$.}
\begin{equation}
    \langle W \rangle = \langle z(\tau_2)\bar z(\tau_1) \rangle\,,
\end{equation}
where the correlation function is computed with the effective action
\begin{equation}
\label{eq:1daction}
    S_{eff} = S_{ABJM} + \int d\tau \Tr (\bar z \cD_{\tau} z)\,.
\end{equation}
Here, the $\tau$-integration is along the Wilson loop contour and the covariant derivative is defined as $\cD_{\tau} = \partial_{\tau}+ i {\cal L}(\tau)$, where ${\cal L}(\tau)$ is the connection of the Wilson operator under investigation. Since we are interested in studying the ordinary Wilson loop, we set $\cD_{\tau} = \partial_{\tau}+ i A_{\mu}\dot x^{\mu}$.

\begin{figure}[ht]
    \centering \includegraphics[width=.2\textwidth]{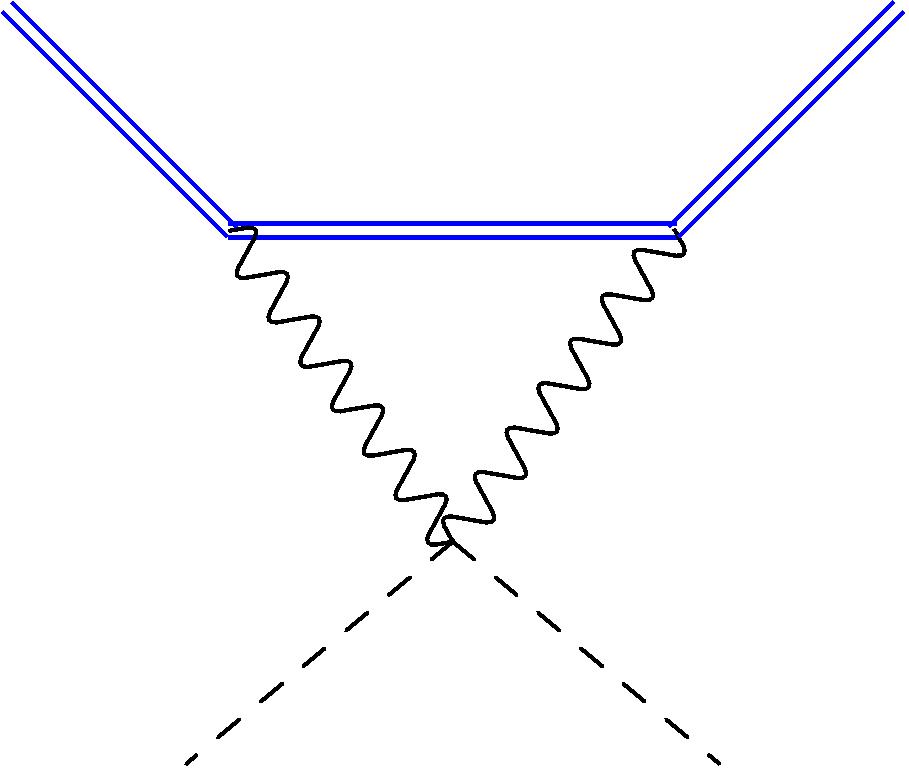}
    \caption{The $\bar z C_I \bar C^I z$ vertex arising at one-loop. Dashed lines correspond to ABJM scalars while double blue straight lines corresponds to the one-dimensional fermion $z$. Wavy lines are gauge fields.}
\label{fig:diagram1}
\end{figure}
The renormalization properties of this operator can be studied by looking at the renormalization of the one-dimensional QFT defined by the action in \eqref{eq:1daction}. 
As shown in \cite{Castiglioni:2022yes}, using the Feynman rules in appendix \ref{app:ren}, at one loop the $\bar z \dot{z}$ kinetic term and the $\bar z A_{\mu}\dot x^{\mu} z$ vertex do not renormalize. Instead, a divergent term arises from the diagram in figure \ref{fig:diagram1}, which corresponds to a new $\bar z C_I \bar C^I z$ vertex, not present in the original action \eqref{eq:1daction}. Its explicit expression is given by
\begin{equation}
\begin{split}
\Gamma^{(\ref{fig:diagram1})}&=\int d\tau_1 \int \tau_2 \int d^d x\, \big(i\bar z A_{\mu}\dot x^{\mu}z\big)(x_1)\big(i\bar z A_{\nu}\dot x^{\nu}z\big)(x_2)\big( A_{\rho} C_I \bar C^I A^{\rho} \big)(x) \\ &= \frac{g^2N}{8\pi\epsilon}\int d\tau\, \bar z C_I\bar C^I z\,.
\end{split}
\end{equation}
The appearance of this UV divergent term spoils renormalizability of the one-dimensional auxiliary theory, unless we include such a new term already at the classical level. This amounts to modifying the covariant derivative in \eqref{eq:1daction} as
$\cD_{\tau} = \partial_{\tau}+ i \left(A_{\mu}\dot x^{\mu} \mp \tfrac{2\pi i}{k}|\dot x|C_I\bar C^I \right)$, which in turn amounts to stating that the correct WLs to consider are 
\begin{equation}
\label{eq:ordWL}
    W^{\pm} = \frac{1}{N} \Tr \cP \exp \left[-i\oint d\tau \left( A_{\mu}\dot x^{\mu} - \tfrac{2\pi i}{k}|\dot x|(\pm\delta_I^J)C_J\bar C^I \right)\right]\,.
\end{equation}
In principle, there are no restrictions on the choice of the sign in front of $\delta_I^J$. We have denoted the two possible options as $W^\pm$. Since there is no field redefinition in the path integral that can compensate for this sign, $W^+$ and $W^-$ are genuine different operators. In the next sections we will further discuss the implications associated with the choice of this sign. 

The scalar couplings in \eqref{eq:ordWL}
preserve the whole $SU(4)$ R-symmetry group, thus the two $W^{\pm}$ operators also do. This implies that in the ABJM theory these are the ``ordinary" maximally R-symmetric WLs. 

This  result is simply the manifestation of the general property according to which any operator preserving the symmetries of the conformal theory at the fixed point gets turned on along the RG flow. 
We emphasize that in the present case this is triggered by a bulk-defect interaction. In fact, the ABJM vertex $A_{\mu} C_I \bar{C}^I A^{\mu}$ plays a central role in constructing the diagram of figure \ref{fig:diagram1}.
As already mentioned, a similar pattern does not arise in $\cN=4$ SYM where, for dimensional reasons, any scalar coupling in the WL must be linear, thus necessarily breaking $SO(6)$. 

Finally, we mention that $W^\pm$ does not preserve any supersymmetry. In fact, supersymmetry invariance requires the scalar coupling matrix $M$ of bosonic loops (see its definition in appendix \ref{app:BPSWL})  to be traceless \cite{Drukker:2022bff}. This is clearly not the case for these operators, for which the scalar coupling matrix is simply $M = \pm {\mathbb{1}}$.

\section{Bosonic flows}
\label{sec:bosonic}

We now begin the study of RG flows which involve fixed points corresponding to the ordinary Wilson loops defined above. We choose to perturb around $W^-$ because, as it will become evident, this corresponds to a UV unstable fixed point. As the simplest case, we consider perturbing with a bosonic operator. 

Referring to \eqref{eq:ordWL}, we consider a four-parameter deformation of $W^-$
\beq \label{eqn:WLsu(4)}
 W(\zeta) =\frac{1}{N} \Tr\cP\exp\bigg[-i\oint d\tau \left( A_{\mu} \dot x^{\mu} + \frac{2\pi i}{k} C_I \bar C^I- \frac{2\pi i}{k}\Delta M_I^{\ J}(\zeta_i)C_J \bar C^I \right)\bigg],
\eeq
where
\beq
\label{eq:Mmatrix3}
 \Delta M(\zeta_i)=2 \begin{pmatrix}
    \zeta_1 & 0 & 0 & 0 \\ 0 & \zeta_2 & 0 & 0\\ 0 & 0 & \zeta_3 & 0 \\ 0 &  0 & 0 & \zeta_4
    \end{pmatrix}\,.
\eeq
For generic values of the parameters this perturbation breaks the $SU(4)$ R-symmetry completely.

According to the calculation presented in appendix \ref{app:ren}, at one loop the  $\beta$-functions parameterising this flow are given by
\begin{equation}\label{eq:betageneric}
    \beta_{\zeta_i} = \frac{g^2N}{2\pi}\zeta_i(\zeta_i-1)\,,\qquad i=1,2,3,4,
\end{equation}
where $g^2=2\pi/k$.
Since the RG equations are decoupled, for the running coupling constants $\zeta_i$ we find 
\beq\label{eq:sol2}
\zeta_i(\mu)=\frac{1}{1+e^{c_i}\mu^{\frac{g^2N}{2\pi}}}\,, \qquad i=1,2,3,4,
\eeq
where $c_i$ are arbitrary real constants. 
The fixed points can be classified and denoted as follows:

\begin{enumerate}
    \item[]\begin{tikzpicture}
  \fill[black] (0,0) rectangle (0.23,0.23);
  \end{tikzpicture}
  \, $SU(4)$ invariant fixed points. The origin $\zeta_i =0, \ i=1,2,3,4,$ trivially corresponds to the ordinary $W^-$ Wilson loop. Similarly, the point $\zeta_i =1, \ i=1,2,3,4,$ corresponds to the ordinary $W^+$ operator. At these two fixed points $SU(4)$ R-symmetry is restored.
    \item[] \begin{tikzpicture}
  \draw[black,fill=black] (0,0) -- 
  ++(0.25,0) --
  ++(-0.125,0.25) -- cycle;
  \end{tikzpicture}
  \, $SU(3)$ invariant fixed points. A first set is obtained by setting one $\zeta_i$ equal to zero and the others equal to one. They correspond to four equivalent Wilson loops with $M = {\text{diag}}(-1,1,1,1)$ and permutations, which exhibit a $SU(3)$ R-symmetry invariance and no supersymmetry. A second set is obtained by setting three $\zeta_i$ equal to zero and one equal to one. In this case the corresponding scalar matrix is $M = {\text{diag}}(1,-1,-1,-1)$ and permutations. Again, the WL is $SU(3)$ invariant, but no supersymmetry is preserved.
    \item[]\begin{tikzpicture}
  \filldraw [blue] (0,0) circle (3pt);
  \end{tikzpicture}
  \, $SU(2) \times SU(2)$ invariant fixed points. These are obtained by setting two $\zeta_i$ equal to zero and two equal to one. They correspond to six equivalent 1/6 BPS bosonic Wilson loops defined in \eqref{eq:WLs}. 
\end{enumerate} 

We can study the RG flows among these fixed points by looking at line, surface and volume projections of the four-dimensional parameter space. In all our pictures arrows go from the UV to the IR. 

\begin{figure}[ht]
    \centering
    \includegraphics[width=0.6\textwidth]{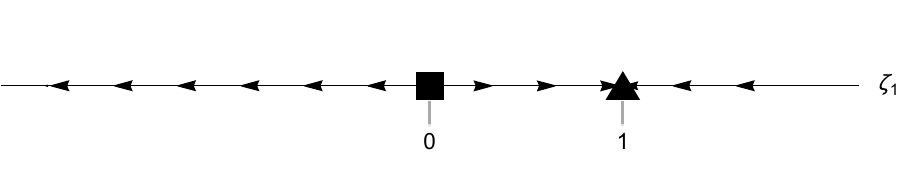}   
    \caption{The RG flow along the $\zeta_1$-line. Arrows go from the UV to the IR. The black square represents $W^-$, whereas the triangle corresponds to a $SU(3)$ invariant bosonic Wilson loop.}
    \label{fig:1bosflow}
\end{figure}

To begin with, we consider turning on only the $\zeta_1$ deformation. This leads to the one-dimensional RG flow depicted in figure \ref{fig:1bosflow}. 
Already in this one-dimensional projection the UV instability of the $W^-$ operator emerges clearly.
 Instead, the point $\zeta_1=1$ corresponding to one of the $SU(3)$ invariant fixed points appears as attractive along the line. 

\begin{figure}[ht]
    \centering
    \includegraphics[width=0.5\textwidth]{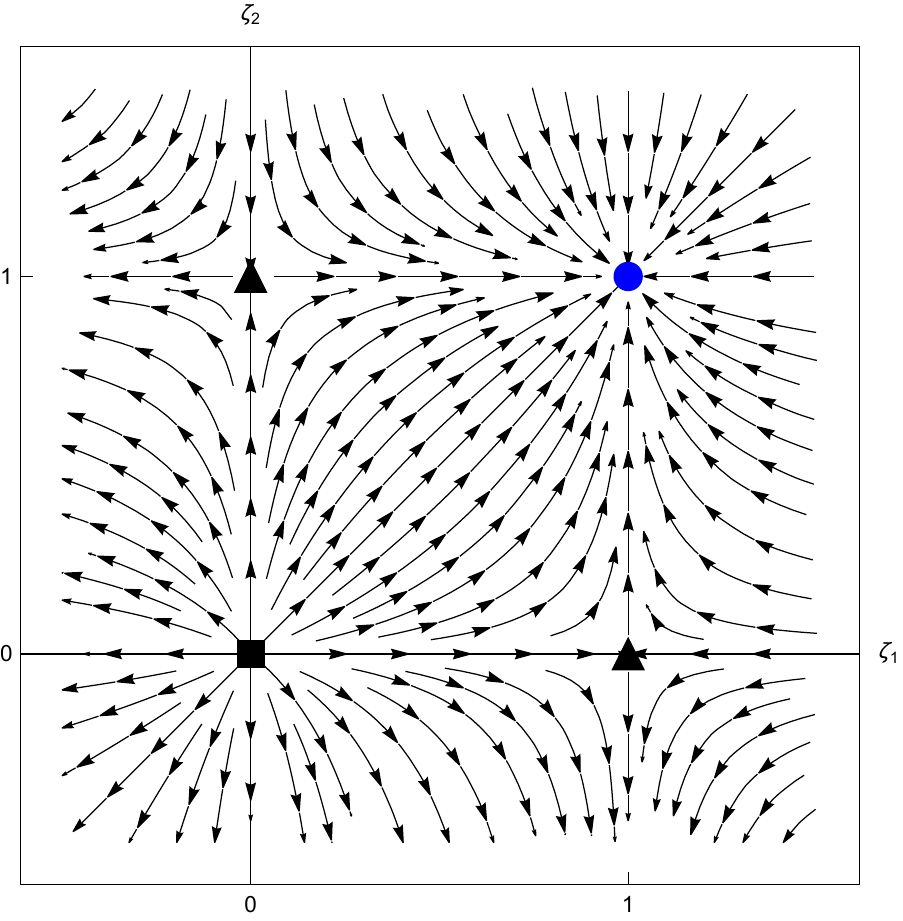}   
    \caption{The RG flow in the $(\zeta_1,\zeta_2)$ plane corresponding to $\zeta_3=\zeta_4=0$. The blue dot is the bosonic $1/6$ BPS Wilson loop, while the two black triangles correspond to bosonic $SU(3)$ invariant Wilson loops.}
    \label{fig:2bosflow}
\end{figure}

We now move to a two-dimensional  section of the parameter space by turning on also the $\zeta_2$ perturbation. The corresponding flows are presented in figure \ref{fig:2bosflow}. We see that under the $\zeta_2$ perturbation the $SU(3)$ fixed point corresponding to the $(1,0)$ point becomes unstable and the RG flow drives the system towards  the point $(1,1)$ marked in blue. This is  the $1/6$ BPS bosonic loop corresponding to $M ={\text{diag}}(1,1,-1,-1)$. 

\begin{figure}[ht]
    \centering
    \includegraphics[width=0.4\textwidth]{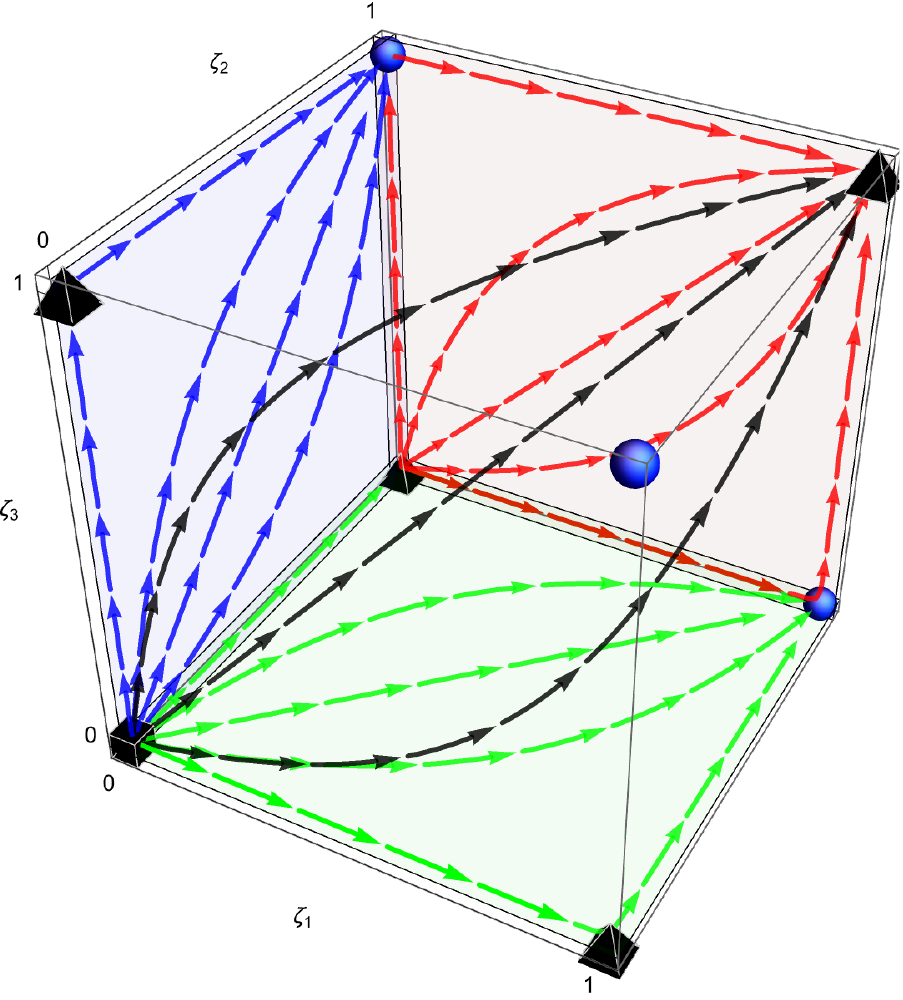}
    \caption{The RG flow in the $(\zeta_1,\zeta_2,\zeta_3)$ space, with $\zeta_4=0$. Blue spheres describe 1/6 BPS bosonic Wilson loops, the black pyramids correspond to $SU(3)$ invariant operators.}
    \label{fig:3bosflow}
\end{figure}

Going one step further, we turn on for instance $\zeta_3$, still keeping $\zeta_4=0$. The resulting RG flows are now depicted in three dimensions, see figure \ref{fig:3bosflow}. We find that as soon as we leave the $\zeta_3=0$ plane, the system is driven towards the $SU(3)$ invariant $(1,1,1)$ point. Moreover, in addition to the $1/6$ BPS operator at $(1,1,0)$ already marked in figure \ref{fig:2bosflow}, two other $1/6$ BPS bosonic loops appear, which correspond to the blue dots at $(1,0,1)$ and $(0,1,1)$. They are simply R-symmetry rotations of each other. The pattern of the flows in figure \ref{fig:2bosflow} is nothing but the projection of the three-dimensional flow on one of the three planes containing the origin. 

Finally, we turn on also the $\zeta_4$ deformation, thus disclosing the whole spectrum of fixed points. In particular, the ordinary Wilson loop $W^+$ now appears, which turns out to be an IR stable fixed point in any direction. The other 14 fixed points, eight $SU(3)$ invariant bosonic WLs and six $1/6$ BPS bosonic operators, are all saddle points and  correspond to the vertices of a four-dimensional hypercube with side 1.

We have considered only the portion of the parameter space within the interval $[0,1]$, because this should correspond to the isolated invariant set \cite{Gukov:2016tnp,10.2307/1995770} of the RG space. Had we considered values outside this interval, we would flow to infinity along runaway directions. This is analogue to what has been found in ${\cal N}=4$ SYM \cite{Polchinski_2011}.


\subsection*{$SU(2)\times SU(2)$ flows}
\label{sec:bosonicSU2}

One interesting subset of deformations corresponds to $SU(2)\times SU(2)$ preserving $\Delta M$ matrices. This necessarily restricts the spectrum of fixed points to a subset of $SU(2) \times SU(2)$ invariant CFTs.

\begin{figure}[ht]
    \centering
    \includegraphics[width=0.5\textwidth]{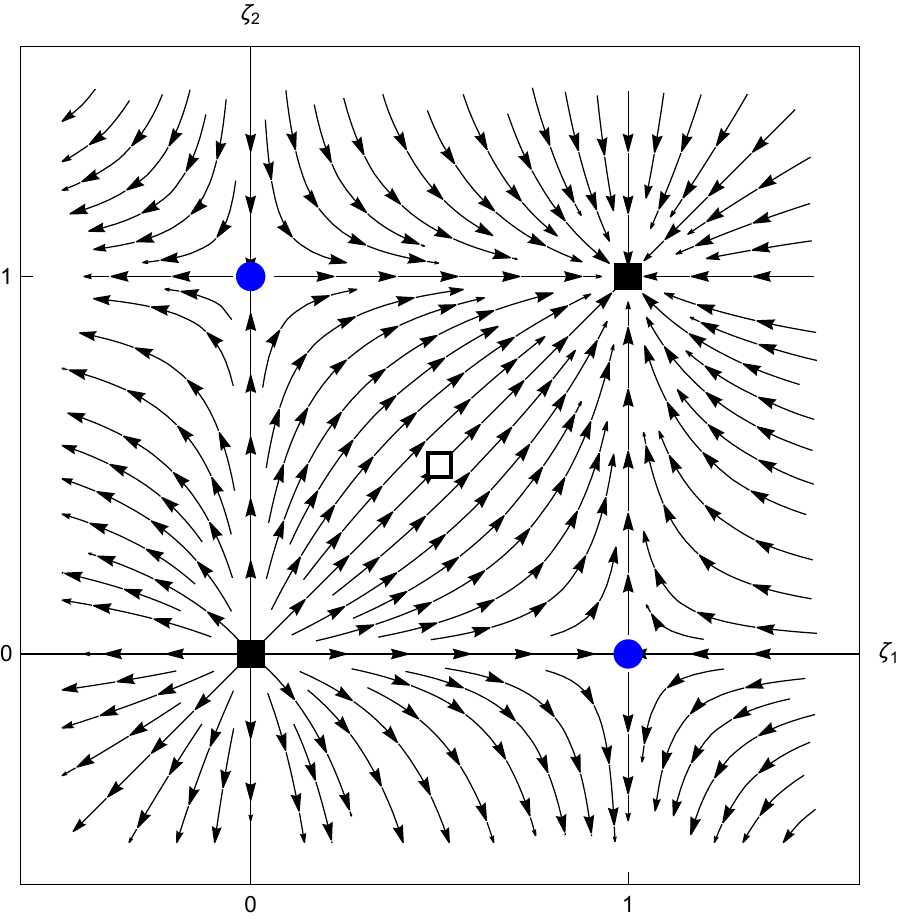}
    \caption{RG flows in the $(\zeta_1,\zeta_2)$ plane. The blue dots correspond to the $SU(2)\times SU(2)$ $1/6$ BPS bosonic loops fixed points, while the black squares correspond to the ordinary $SU(4)$ invariant Wilson loops. The white square is the Wilson loop with purely gauge connection $A_{\mu}\dot x^{\mu}$.}
    \label{fig:RGflow4}
\end{figure}

To this end we consider perturbing $W^-$ as in \eqref{eqn:WLsu(4)} with
\beq
\label{eq:Mmatrix2}
\Delta M(\zeta_1,\zeta_2)=2 \begin{pmatrix}
    \zeta_1 & 0 & 0 & 0 \\ 0 & \zeta_1 & 0 & 0\\ 0 & 0 & \zeta_2 & 0 \\ 0 &  0 & 0 & \zeta_2
    \end{pmatrix}\,.
\eeq
Such a deformation gives rise to a two-parameter family of interpolating Wilson loops. The one-loop $\beta$-functions and their solutions can be easily read from \eqref{eq:betageneric} and \eqref{eq:sol2} and the corresponding flows are depicted in figure \ref{fig:RGflow4}. They  connect $W^-$ (for $\zeta_1 = \zeta_2 = 0$), $W^+$ (for $\zeta_1 = \zeta_2 = 1$) and the bosonic BPS Wilson loop $W_{1/6}$ (for $\zeta_1 = 1, \, \zeta_2 = 0$ or $\zeta_1 = 0, \, \zeta_2 = 1$). The $W(\zeta_1,\zeta_2)$ operators possess a $\mathbb{Z}_2$ symmetry under the $\zeta_1 \leftrightarrow \zeta_2$ exchange, which comes from the freedom to exchange the two $SU(2)$ factors in the R-symmetry group. This symmetry is also manifest in the behavior of the RG trajectories. 

We note that although the deformation is not the same and thus fixed points are different, the flows in figure \ref{fig:2bosflow} and \ref{fig:RGflow4} are equal. In fact, they both share the same behavior as the RG flow of the $O(N)$ model (see for instance figures 4 and 7 of \cite{Gukov:2016tnp}).


\subsection*{$SU(4)$ flows}
\label{sec:bosonicSU4}

A further interesting subset of flows is the one induced by a single parameter deformation $\Delta M_I^{\ J} (\zeta) = 2 \zeta \, \delta_I^{J}$. In the same spirit as $\cN=4$ SYM, this deformation produces a one-parameter family of interpolating Wilson loops, which connects $W^-$ for $\zeta=0$ and $W^+$ for $\zeta=1$.

\begin{figure}[ht]
    \centering
    \includegraphics[width=0.6\textwidth]{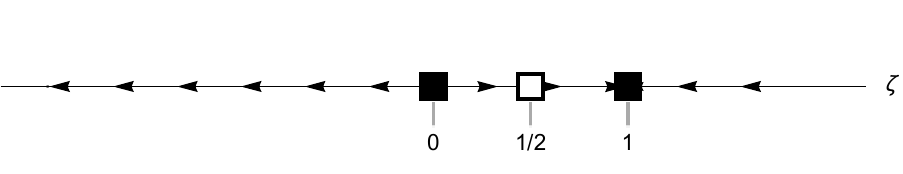}   
    \caption{The RG flow along the $\zeta$-line. The black squares represent $W^-$ and $W^+$ at $\zeta=0$ and $\zeta=1$, respectively. The white square at $\zeta=1/2$ represents the Wilson loop with connection $A_\mu\dot{x}^\mu$. }
    \label{fig:su(4)flow}
\end{figure}

As is evident from figure \ref{fig:su(4)flow}, the ordinary Wilson loop $W^{-}$ corresponding to a negative scalar coupling $M = - {\mathbb 1}$ is a UV unstable fixed point. On the other hand, the ordinary loop $W^+$ with $M = + {\mathbb 1}$ is an IR stable point. We note that the naive analogue of the ordinary Wilson loop in $\cN=4$ SYM, \textit{i.e.} the Wilson loop corresponding to a pure gauge connection $A_{\mu}\dot x^{\mu}$, is located at point $\zeta=1/2$. Consistently with our previous findings, it does not correspond to any fixed point, rather it belongs to the RG straight line that connects the two ``true'' ordinary WLs, $W^-$ and $W^+$.

The main conclusion we can draw from the present investigation is that at quantum level the $W^+$ and $W^-$ operators exhibit a very different behavior under renormalization, though classically they simply differ by the overall sign in front of the scalar coupling. More evidence on their deep different nature will emerge in the next sections.

\section{Fermionic flows}\label{sec:fer1}

In three-dimensional Chern-Simons-matter theories - ABJM theory included - a more general class of Wilson loops can be considered, which involves coupling to fermions \cite{Drukker:2009hy,Cardinali:2012ru}. In this case the connection appearing in the exponent of the Wilson operator is promoted to a supermatrix $\cL$ where, in addition to gauge fields and scalar bilinears in the diagonal elements, fermi fields appear linearly in the off-diagonal entries\footnote{For longer quiver gauge theories, this construction can be generalized to build loops which couple to more than two nodes. See for instance \cite{Mauri_2017,Drukker:2020dvr,Drukker:2022txy}.}.

In the same spirit as in the previous section, we can study flows driven by adding fermionic deformations. As a first case, here we will keep the diagonal structure of $\cL$ fixed, \textit{i.e.} gauge fields and scalar bilinears with a fixed structure, and discuss possible flows obtained by simply adding fermionic fields to the off-diagonal entries of $\cL$. 

We start with a composite non-BPS bosonic loop with superconnection
\begin{equation}\label{eq:superconnection}
    \cL = \begin{pmatrix}
    A_{\mu} \dot{x}^\mu - \frac{2\pi i}{k}\vert\dot{x}\vert M_I^{\ J}C_J \bar C^I & 0 \\
    0 & \hat A_{\mu} \dot{x}^\mu - \frac{2\pi i}{k}\vert\dot{x}\vert M_I^{\ J} \bar C^I C_J 
    \end{pmatrix},
\end{equation}
where $A,\hat A$ are charged under the two nodes of the ABJM theory and $M$ is a fixed - not better specified -  scalar coupling matrix. We take the fermionic perturbation to be of the following form
\begin{equation}\label{eq:ferdef1}
    \Delta \cL_F = -ig\begin{pmatrix}
        0 &  \eta  \left(\chi_1\bar\psi^{1} + \chi_2\bar\psi^{2}+\chi_3\bar\psi^{3}+\chi_4\bar\psi^{4}\right) \\ \left(\chi_{1}\psi_1 + \chi_2\psi_2 +\chi_3\psi_3 +\chi_4\psi_4 \right)\bar \eta & 0
    \end{pmatrix}\,,
\end{equation}
where $\chi_i, \ i=1,2,3,4$, are four arbitrary real parameters, the bosonic spinorial couplings $\eta^{\alpha}, \bar{\eta}_\alpha$ on the circle are defined as in \cite{Cardinali:2012ru}
\begin{equation}\label{eq:etas}
\begin{split}
    & \eta = ie^{\frac{i}{2} \ell \tau} \left[ \cos\left( 
(1-\ell)\tfrac{\pi}{4} \right) - e^{i\tau}\sin\left((1-\ell)\tfrac{\pi}{4}\right)
 \right]\begin{pmatrix} 1 & \, -i \ell e^{-i\tau} \end{pmatrix}\,,\\
    &\bar\eta = i e^{-\frac{i}{2} \ell\tau}\left[ \cos\left( 
(1-\ell)\tfrac{\pi}{4} \right) - e^{-i\tau}\sin\left((1-\ell)\tfrac{\pi}{4}\right) \right] \begin{pmatrix} -i\\  \ell e^{i\tau} \end{pmatrix} \, , 
\end{split}
\end{equation}
and the spinorial products are always meant to be 
$\lambda \bar\rho \equiv \lambda^{\alpha} \bar\rho_{\alpha}$.
The constant parameter $\ell$ in \eqref{eq:etas} can take only the two values $\pm 1$. Therefore, we have two branches of fermionic deformations, which differ by the sign of $\ell$. In principle, the structure of $\eta,\bar\eta$ is totally arbitrary, however we fix them as in \eqref{eq:etas} in order to generate flows that can reach $1/2$ BPS operators (see appendix \ref{app:BPSWL} for their definition).

For arbitrary parameters $\chi_i$, the fermionic deformation in \eqref{eq:ferdef1} breaks completely the R-symmetry group. However, if the scalar matrix $M$ is originally chosen to preserve a subgroup of the R-symmetry, whichever fermions we plan to add to $\cL$, they have to preserve the same R-symmetry structure. In fact, if this were not the case the R-symmetry mismatch between the fermion and scalar couplings would give rise to UV divergent contributions that would necessarily turn on a deformation of $M$ along the flow. In the auxiliary one-dimensional field approach, this is made manifest by the fact that, as explained in appendix \ref{app:fermdef},  already at one loop a divergent triangle fermionic diagram with external fields $\bar z C \bar C z$ arises, see figure \ref{subfig:diagc}, which is proportional to the fermionic deformation. It follows that in order to cancel this divergence we have to add a parametric deformation in the scalar coupling matrix, as well.
This general observation has interesting consequences. 

First of all, since $SU(4)$ invariant fermionic couplings do not exist, it is not possible to deform a $SU(4)$ invariant ordinary WL by adding purely fermionic deformations. A fermionic perturbation of the ordinary $W^\pm$ operators considered above would necessarily require adding also a bosonic perturbation. This kind of perturbations will be discussed in section \ref{sec:generic}. 

Sticking to pure fermionic deformations, the most we can do is to start with a $SU(3)$ invariant superconnection and add a $SU(3)$ invariant fermionic coupling. 
For instance, we can perturb around the bosonic WL corresponding to superconnection \eqref{eq:superconnection} with scalar coupling matrix $M=\tilde\ell\,\text{diag}(- 1, 1, 1,1)$, with $\tilde\ell=\pm 1$ independent of $\ell$ (see definition \eqref{eq:bosonicsu3}). In order to preserve $SU(3)$ R-symmetry along the RG flows, it is crucial to choose in \eqref{eq:ferdef1} only one non-vanishing fermion, precisely the one carrying the same R-symmetry index of the scalar bilinear which appears in  $M C \bar{C}$  with opposite sign. Therefore, given $M=\tilde\ell\,\text{diag}(- 1, 1, 1,1)$, we are forced to take $\chi_1 \neq 0$ and $\chi_2=\chi_3=\chi_4=0$ in \eqref{eq:ferdef1}. If we were to choose a different scalar coupling matrix, for instance $M=\tilde\ell\,\text{diag}(1, -1, 1,1)$, we should perturb with $\chi_2$. 

The $\beta$-functions for single fermion deformations can be computed following \cite{Castiglioni:2022yes}. Turning on the $\chi_i$ deformation and evaluating the corresponding $\beta$-function for generic $\ell$, we find
\begin{equation}\label{eq:beta-functions}
    \beta_{\chi_i} = \frac{g^2N}{2\pi}\ell\left( \chi_i^2-1 \right)\chi_i\,.
\end{equation}
 The effect of $\ell$ on the $\beta$-function is a relevant overall sign. 
 
From \eqref{eq:beta-functions} we see that there is one fixed point at $\chi_i=0$, corresponding to the undeformed operator, \emph{i.e.} the non-BPS bosonic $SU(3)$ Wilson loop, and two other ones at $\chi_i=\pm 1$. However, since the deformed WL is invariant under $\chi_i \to - \chi_i$, without losing generality we restrict our study to $\chi_i\ge 0$.

\begin{figure}[ht]
    \centering
    \subfigure[]{
    \includegraphics[width=0.45\textwidth]{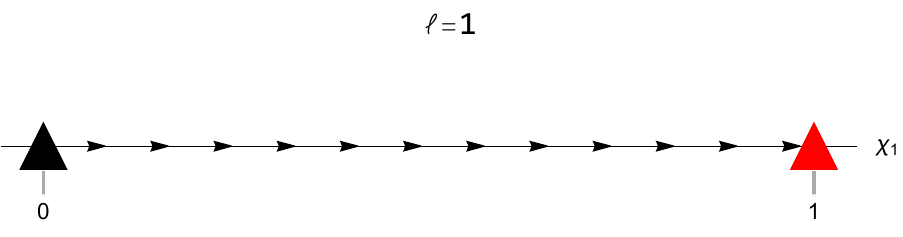}
    \label{subfig:fermflow1}}\quad
    \subfigure[]{
    \includegraphics[width=0.45\textwidth]{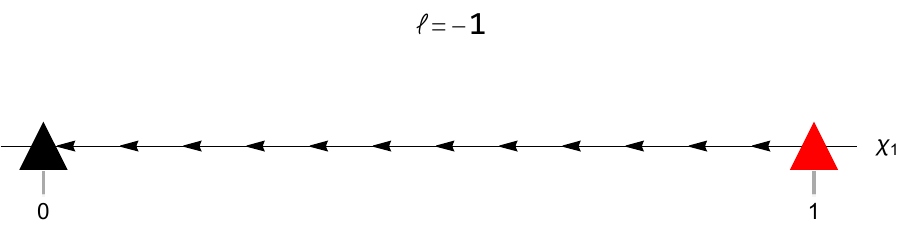}
    \label{subfig:fermflow2}} 
   \caption{$SU(3)$ symmetric fermionic flows for \subref{subfig:fermflow1} $\ell=1$ and \subref{subfig:fermflow2} $\ell=-1$. The black triangles at the origin represent non-BPS bosonic $SU(3)$ operators, whereas the red triangles are $1/2$ BPS Wilson loops. 
   }
    \label{fig:fermflows}
\end{figure}

In the study of these flows we have two different possibilities corresponding to the sign of $\ell=\tilde\ell$ \footnote{When $\ell =-\tilde\ell$ the theory is not renormalizable, since divergent contributions in the bosonic sector arise. This means turning on a mixed bosonic-fermionic flow, as we study in section \ref{sec:generic}.}. We interpolate between $SU(3)$ bosonic operators (\begin{tikzpicture}\draw[black,fill=black] (0,0) -- 
  ++(0.25,0) --
  ++(-0.125,0.25) -- cycle;
  \end{tikzpicture}) and $1/2$ BPS ones (\begin{tikzpicture}\draw[red,fill=red] (0,0) -- 
  ++(0.25,0) --
  ++(-0.125,0.25) -- cycle;
  \end{tikzpicture}), see \eqref{eq:W1/2pm}. In particular, for $\ell=1$ the bosonic operator is unstable and flows to the stable $W_{1/2}^+$. In contrast, for $\ell=-1$ the bosonic operator is stable while $W_{1/2}^-$ is unstable. This pattern is represented in figure \ref{fig:fermflows}.

So far we have considered only single fermion deformations. One might wonder whether a more general pattern exists, where more than one fermion parametric deformation is turned on. The answer is no. In fact, as one can infer from the calculations in appendix \ref{app:fermdef}, as soon as we turn on a second $\chi_j$ coupling without a suitable accompanying bosonic deformation, the one-dimensional auxiliary theory is not renormalizable. In other words,
deforming with more than one fermionic coupling necessarily turns on also a bosonic deformation along the RG flows.
\section{Mixed flows}
\label{sec:generic}

As a last step, we consider moving out from the ordinary $W^-$ fixed point by adding a combination of both bosonic \eqref{eq:Mmatrix3} and fermionic \eqref{eq:ferdef1} deforming operators. In general, the perturbation will depend on eight parameters $\zeta_i, \chi_i$, with $i=1,2,3,4$. 

The $\beta$-functions for the fermionic deformations are the straightforward generalization of the ones evaluated in \eqref{eq:beta-functions}. At the order we are working, they are not affected by the presence of bosonic deformations and read 
\begin{equation}
\label{eq:betachis}
    \beta_{\chi_i} = \frac{g^2N}{2\pi}\left( \ell_1 \chi_1^2 + \ell_2\chi_2^2 + \ell_3\chi_3^2 + \ell_4\chi_4^2  -\ell_i \right)\chi_i \, , \qquad i=1,2,3,4,
\end{equation}
where $\ell_i=\pm 1$ distinguishes between the two possible branches of fermionic deformations, as described in section \ref{sec:fer1}.

Instead, the $\beta$-functions of the scalar couplings $\zeta_i$ get affected by the presence of the fermionic deformations, in fact they acquire a non-trivial dependence on $\chi_i$. To simplify the discussion, 
in the following sections we investigate in details the cases where only one or two fermions appear in the deformation. Turning on more than two fermionic parameters does not add much to the discussion. The results, although more involved, exhibit the same qualitative behavior. 

\subsection{Bosonic plus one-fermion deformations}
\label{sec:bosplus1ferm}

We start discussing RG flows driven by the bosonic deformation $\Delta M(\zeta_i)$ in \eqref{eq:Mmatrix3} plus the fermionic one in \eqref{eq:ferdef1} where we set 
$\chi_2=\chi_3=\chi_4=0$. As long as we are interested in single fermion perturbations, this choice is completely general, as deformations generated by $\chi_2, \chi_3$ or $\chi_4$ can be obtained by simply applying an R-symmetry rotation.

The $\beta$-function for $\chi_1$ can be easily read from \eqref{eq:betachis}. For $\chi_2=\chi_3=\chi_4=0$ it reduces to \eqref{eq:beta-functions} with $i=1$.

The $\beta$-functions for the bosonic deformations are computed in appendix \ref{app:fermdef}. 
More precisely, they can be read from \eqref{eq:betasdoubleferm} setting $\chi_a=\chi_1$ and $\chi_b=0$
\begin{equation}\label{eq:beta2}
\begin{split}
    \beta_{\zeta_1} &= \frac{g^2N}{2\pi}\left( \zeta_1 -1 +\ell \chi_1^2 + \frac{(1-\ell)}{2}\frac{\chi_1^2}{\zeta_1} \right)\zeta_1\,,\\
    \beta_{\zeta_k} & = \frac{g^2N}{2\pi}\left( \zeta_k -1 +\ell \chi_1^2- \frac{(1+\ell)}{2} \frac{\chi_1^2}{\zeta_k} \right)\zeta_k\,, \qquad\qquad k=2,3,4\,.
\end{split}
\end{equation}
These $\beta$-functions together with $\beta_{\chi_1}$ describe RG flows in a five-dimensional space. As already shown in section \ref{sec:fer1}, the non-trivial dependence of $\beta_{\chi_1}$ on $\ell$ leads to two qualitatively different classes of flows. Therefore, we discuss the two cases, $\ell= 1$ and $\ell=-1$, separately. 

\paragraph{$\pmb{\ell=1}$ case.}
In this case the $\beta$-functions reduce to
\begin{equation}\label{eq:beta1fermion}
\begin{split}
    \beta_{\chi_1} &= \frac{g^2N}{2\pi}(\chi_1^2-1)\chi_1\,,\\
    \beta_{\zeta_1} &= \frac{g^2N}{2\pi}\left( \zeta_1 -1 + \chi_1^2 \right)\zeta_1\,,\\
    \beta_{\zeta_k} & = \frac{g^2N}{2\pi}\left( \zeta_k(\zeta_k -1 +\chi_1^2)-\chi_1^2 \right)\,, \qquad\qquad k=2,3,4\,.
\end{split}
\end{equation}

The nature of the RG flows and the stability of the fixed points in the  five dimensional space can be understood by plotting the solutions to the $\beta$-function equations 
$\mu \frac{\partial \Xi}{\partial \mu} = \beta_\Xi$, for any coupling $\Xi = (\chi_1,\zeta_1,\zeta_k)$. The solutions depend on five arbitrary parameters and read explicitly
\begin{equation}
\label{eq:solutionsl1}
\begin{split}
    \chi_1(\mu) & = \frac{1}{\sqrt{e^{2c_0}\mu^2 +1}}\,,\\
    \zeta_1 (\mu) & = \frac{1}{\sqrt{e^{2c_0}\mu^2 +1}}\left(\arctanh \left(\sqrt{e^{2c_0}\mu^2 +1}\right) +c_1\right)^{-1}\,,\\
    \zeta_k (\mu) & = 1 - \frac{\mu^2}{e^{-2c_0}+\mu^2-c_k\sqrt{1+e^{2c_0}\mu^2}}\,.
\end{split}
\end{equation}

\begin{figure}[ht]
    \centering
    \subfigure[]{
    \includegraphics[width=0.31\textwidth]{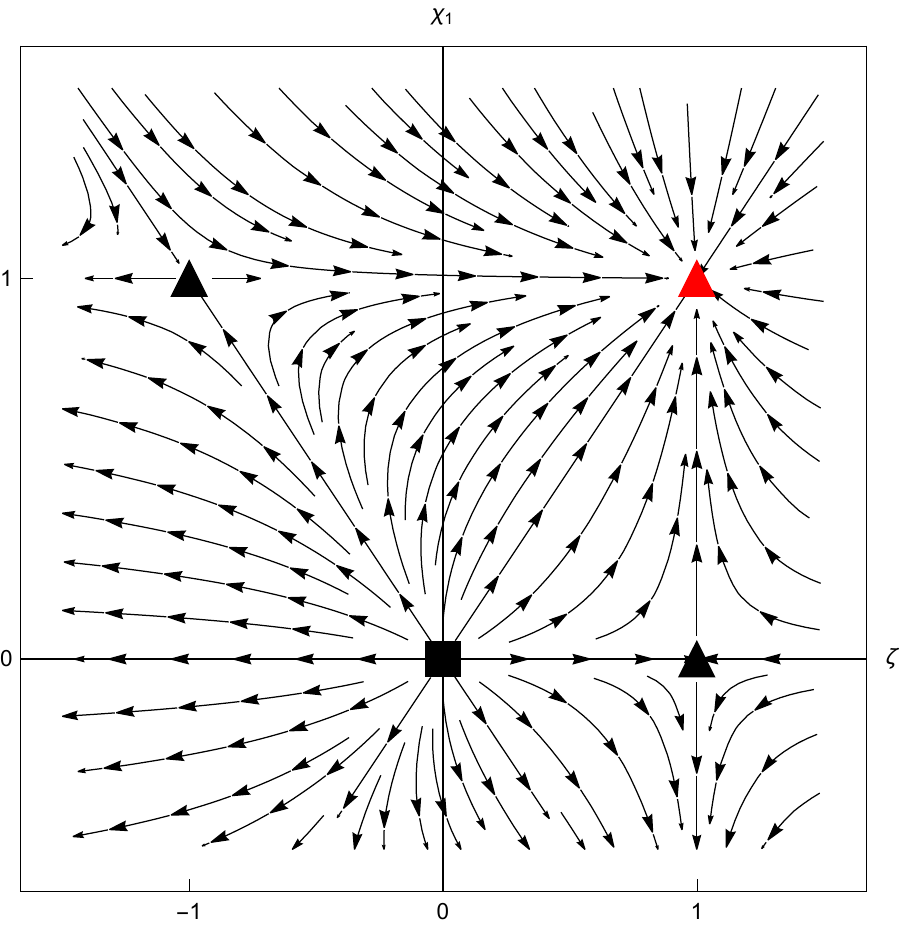}
    \label{subfig:1fermflow1}}
    \subfigure[]{
    \includegraphics[width=0.31\textwidth]{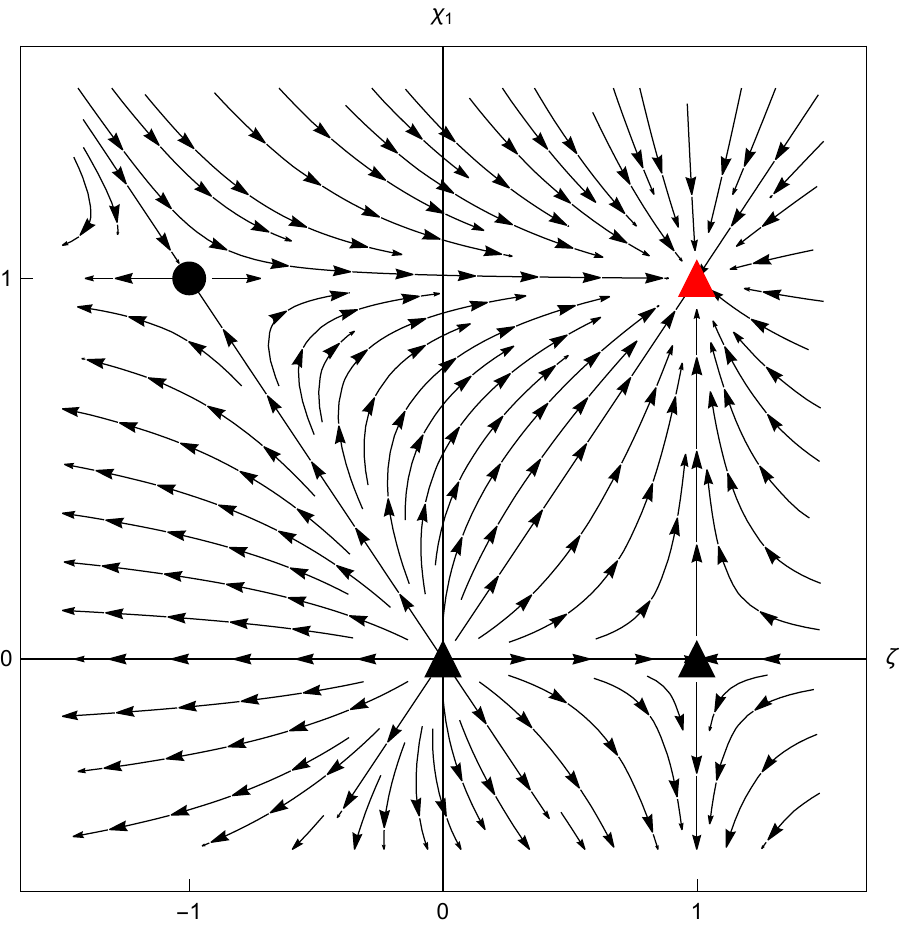}
    \label{subfig:1fermflow2}}  
    \subfigure[]{
    \includegraphics[width=0.31\textwidth]{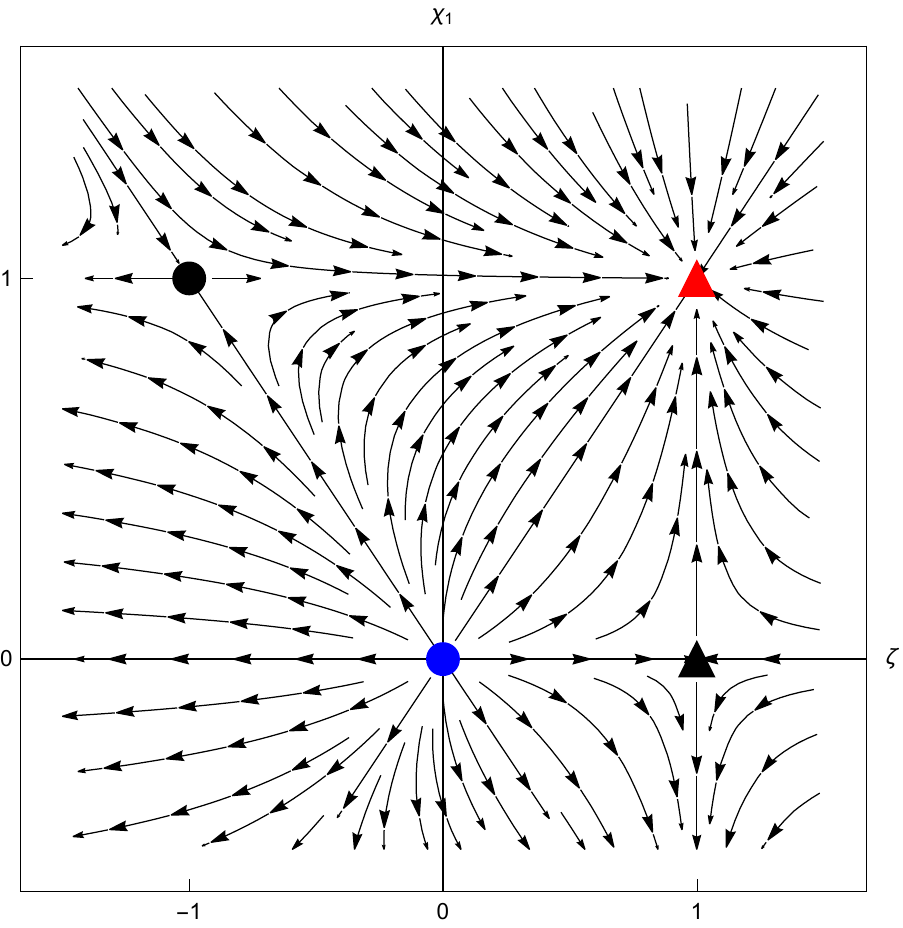}
    \label{subfig:1fermflow3}}
    \caption{The $\ell = 1$ RG flows \subref{subfig:1fermflow1} in the  $(\zeta_1=0, \zeta_2=\zeta_3=\zeta_4 \equiv \zeta, \chi_1)$ plane, \subref{subfig:1fermflow2} in the  $(\zeta_1=0, \zeta_2= 1, \zeta_3=\zeta_4 \equiv \zeta, \chi_1)$ plane and \subref{subfig:1fermflow3} in the $(\zeta_1=0, \zeta_2\equiv\zeta, \zeta_3=\zeta_4=1, \chi_1)$ plane. All cases have similar flows but with fixed points corresponding to different operators.}
    \label{fig:1fermflow}
\end{figure}

In order to have a visual description of the RG flows, in figures \ref{fig:1fermflow} we provide projections on a few planes where fixed points can be detected. To this end, we note that as long as $\chi_1\ne 0$ - which is our case of interest - the last term in $\beta_{\zeta_k}$ prevents from projecting on planes where $\zeta_k=0$, $k=2,3,4$. We then choose to project on $\zeta_1=0$ setting either $\zeta_2=\zeta_3=\zeta_4\equiv\zeta$ (figure \ref{subfig:1fermflow1})  or $\zeta_2=1$, $\zeta_3=\zeta_4 \equiv \zeta$ (figure \ref{subfig:1fermflow2})
or $\zeta_3=\zeta_4=1$, $\zeta_2\equiv\zeta$ (figure \ref{subfig:1fermflow3}).

In figure \ref{subfig:1fermflow1} we recognize the $W^-$ unstable point at the origin (\begin{tikzpicture}\draw[black,fill=black] (0,0) -- 
  ++(0.23,0) --
  ++(0,0.23) -- ++(-0.23,0) -- cycle;
  \end{tikzpicture}) and the IR stable fixed point $W^+_{1/2}$ (\begin{tikzpicture}\draw[red,fill=red] (0,0) -- 
  ++(0.25,0) --
  ++(-0.125,0.25) -- cycle;
  \end{tikzpicture}). The other two saddle points correspond to non-BPS $SU(3)$ invariant operators. The point $(\zeta,\chi_1)=(1,0)$ is a bosonic WL with scalar coupling matrix $M=\text{diag}(-1,1,1,1)$ (\begin{tikzpicture}\draw[black,fill=black] (0,0) -- 
  ++(0.25,0) --
  ++(-0.125,0.25) -- cycle;
  \end{tikzpicture}), whereas $(\zeta,\chi_1)=(-1,1)$ corresponds to a fermionic WL with $M=\text{diag}(-1,-3,-3,-3)$ and non-BPS coupling to $\psi_1,\bar\psi^1$ (still indicated with \begin{tikzpicture}\draw[black,fill=black] (0,0) -- 
  ++(0.25,0) --
  ++(-0.125,0.25) -- cycle;
  \end{tikzpicture}).

In figure \ref{subfig:1fermflow2} the origin and the point $(\zeta,\chi_1)=(1,0)$ correspond to non-BPS SU(3) invariant bosonic WL (\begin{tikzpicture}\draw[black,fill=black] (0,0) -- 
  ++(0.25,0) --
  ++(-0.125,0.25) -- cycle;
  \end{tikzpicture}). The flow still reaches the fermionic $1/2$ BPS $W_{1/2}^+$ (\begin{tikzpicture}\draw[red,fill=red] (0,0) -- 
  ++(0.25,0) --
  ++(-0.125,0.25) -- cycle;
  \end{tikzpicture}) and a novel non-BPS SU(2) invariant fermionic operator with $M=\text{diag}(-1,1,-3,-3)$ and $\psi_1,\bar\psi^1$ couplings (\begin{tikzpicture}
  \filldraw [black] (0,0) circle (3pt);
  \end{tikzpicture}).

In figure \ref{subfig:1fermflow3} the flows connect the $1/6$ BPS bosonic WL (\begin{tikzpicture}
  \filldraw [blue] (0,0) circle (3pt);
  \end{tikzpicture}) at the origin with a non-BPS SU(3) invariant bosonic point (\begin{tikzpicture}\draw[black,fill=black] (0,0) -- 
  ++(0.25,0) --
  ++(-0.125,0.25) -- cycle;
  \end{tikzpicture}), the fermionic $1/2$ BPS $W^+_{1/2}$ (\begin{tikzpicture}\draw[red,fill=red] (0,0) -- 
  ++(0.25,0) --
  ++(-0.125,0.25) -- cycle;
  \end{tikzpicture}) and a novel non-BPS $SU(2)$ invariant fermionic operator with $M=\text{diag}(-1,-3,1,1)$ and $\psi_1,\bar\psi^1$ couplings (\begin{tikzpicture}
  \filldraw [black] (0,0) circle (3pt);
  \end{tikzpicture}). 

\begin{figure}[ht]
    \centering
    \includegraphics[width=0.5\textwidth]{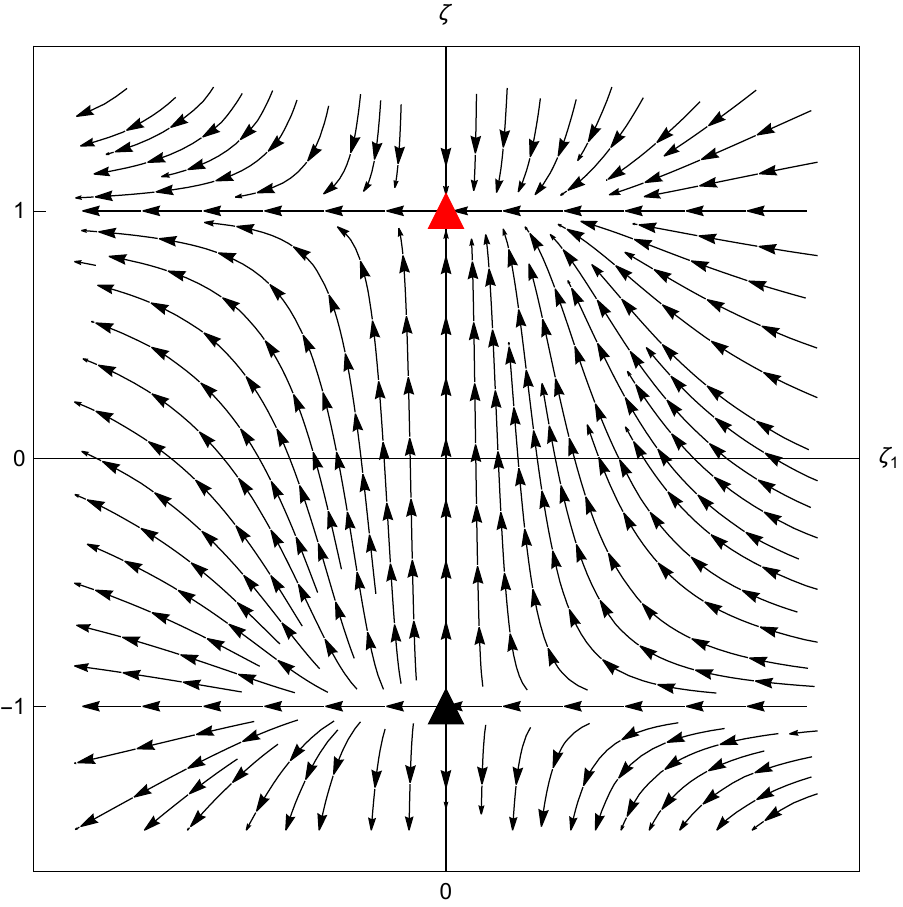}
    \caption{Section $\chi_1=1$ in the $(\zeta_1,\zeta)$ plane for $\zeta_2=\zeta_3=\zeta_4 \equiv \zeta$.}
    \label{fig:test}
\end{figure}

As a further case, in figure \ref{fig:test} we consider the $\chi_1=1$ section, taking $\zeta_1$ generically non-vanishing and setting $\zeta_2=\zeta_3=\zeta_4\equiv \zeta$. We see that it exhibits a flow between the non-BPS SU(3) invariant fermionic operator with $M=\text{diag}(-1,-3,-3,-3)$ (\begin{tikzpicture}\draw[black,fill=black] (0,0) -- 
  ++(0.25,0) --
  ++(-0.125,0.25) -- cycle;
  \end{tikzpicture}) and the $1/2$ BPS WL $W^+_{1/2}$ (\begin{tikzpicture}\draw[red,fill=red] (0,0) -- 
  ++(0.25,0) --
  ++(-0.125,0.25) -- cycle;
  \end{tikzpicture}).

\paragraph{$\pmb{\ell=-1}$ case.} A similar analysis can be done for $\ell = - 1$. In this case the $\beta$-functions in (\ref{eq:beta-functions}, \ref{eq:beta2}) reduce to
\begin{equation}
\begin{split}
    \beta_{\chi_1} &= -\frac{g^2N}{2\pi}(\chi_1^2-1)\chi_1\,,\\
    \beta_{\zeta_1} &= \frac{g^2N}{2\pi}\left( \zeta_1( \zeta_1 -1 -\chi_1^2) + \chi_1^2 \right)\,,\\
    \beta_{\zeta_k} & = \frac{g^2N}{2\pi}\left( \zeta_k -1 - \chi_1^2\right)\zeta_k\,, \qquad\qquad k=2,3,4\,.
\end{split}
\end{equation}
and solving the corresponding flow equations we find
\begin{equation}
\label{eq:solutionsl-1}
\begin{split}
    \chi_1(\mu) & = \frac{\mu}{\sqrt{e^{2c_0} +\mu^2}}\,,\\
    \zeta_1 (\mu) & = 1- \frac{\mu}{\sqrt{e^{2c_0} +\mu^2}}\left(\arctanh \left(\frac{\mu}{\sqrt{e^{2c_0}+\mu^2}}\right) -c_1\right)^{-1}\,,\\
    \zeta_k (\mu) & = \frac{1}{e^{-2c_0}\mu^2+1+ c_k \mu\sqrt{e^{2c_0}+\mu^2}}\,.
\end{split}
\end{equation}
The main differences with the previous case are the crucial sign change in  $\beta_{\chi_1}$ and somehow the exchange of the roles of $\zeta_1$ with $\zeta_k$. In particular, in order to project on planes containing fixed points, this time $\zeta_1=0$ is not a consistent choice, whereas we can project on $\zeta_k=0, (k=2,3,4)$ hyperplanes. 

\begin{figure}[H]
    \centering
    \subfigure[]{
    \includegraphics[width=0.4\textwidth]{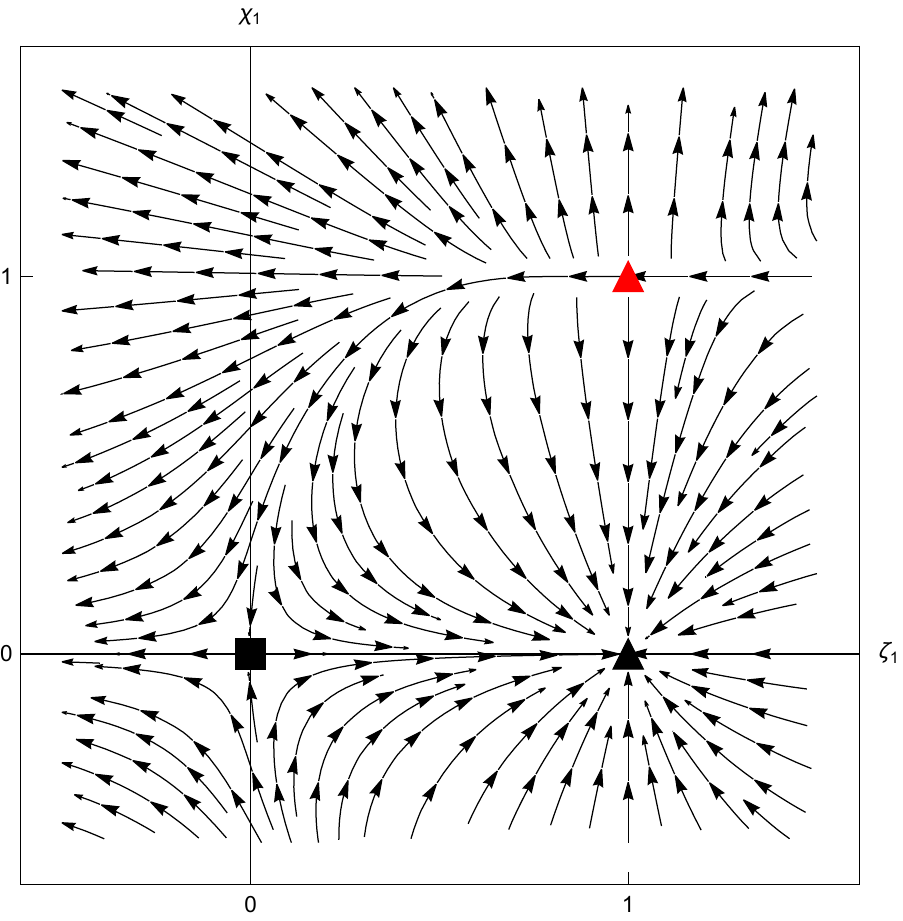}
    \label{subfig:2fermflow1}}
    \subfigure[]{
    \includegraphics[width=0.4\textwidth]{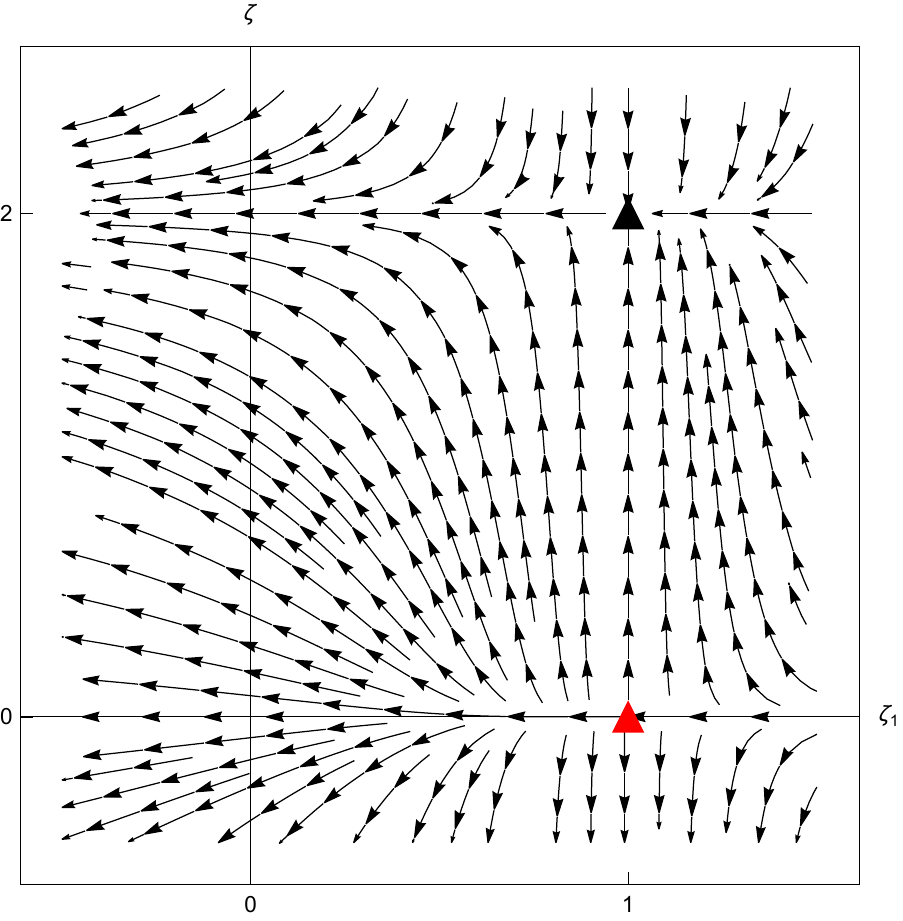}
    \label{subfig:2fermflow2}}   
    \caption{The $\ell = -1$ RG flows \subref{subfig:1fermflow1} in the $(\zeta_1\equiv \zeta,\chi_1)$ plane with $\zeta_2 = \zeta_3 = \zeta_4 =0$ and \subref{subfig:1fermflow2} in the  $(\zeta_1,\zeta)$ plane with $\chi_1=1$ and $\zeta_2=\zeta_3=\zeta_4\equiv\zeta$.}\label{fig:2fermflow}
\end{figure}

In figure \ref{subfig:2fermflow1} we plot the flows in the $(\zeta_1 \equiv \zeta, \chi_1)$ plane, setting $\zeta_2 = \zeta_3 = \zeta_4 =0$, whereas figure \ref{subfig:2fermflow2} describes the flows in the $\chi_1 =1$ section, taking $\zeta_1$ and $\zeta_2 = \zeta_3 = \zeta_4 \equiv \zeta$ free. The pattern is similar to the $\ell=1$ case, the main difference being that the red triangle now corresponds to the 1/2 BPS fermionic loop $W^-_{1/2}$. Moreover, the black triangle in the second figure corresponds to a $SU(3)$ invariant non-BPS fermionic operator with scalar coupling $M=\text{diag}(1,3,3,3)$. 

\vskip 10pt

\begin{figure}[ht]
    \centering
    \subfigure[]{
    \includegraphics[width=0.40\textwidth]{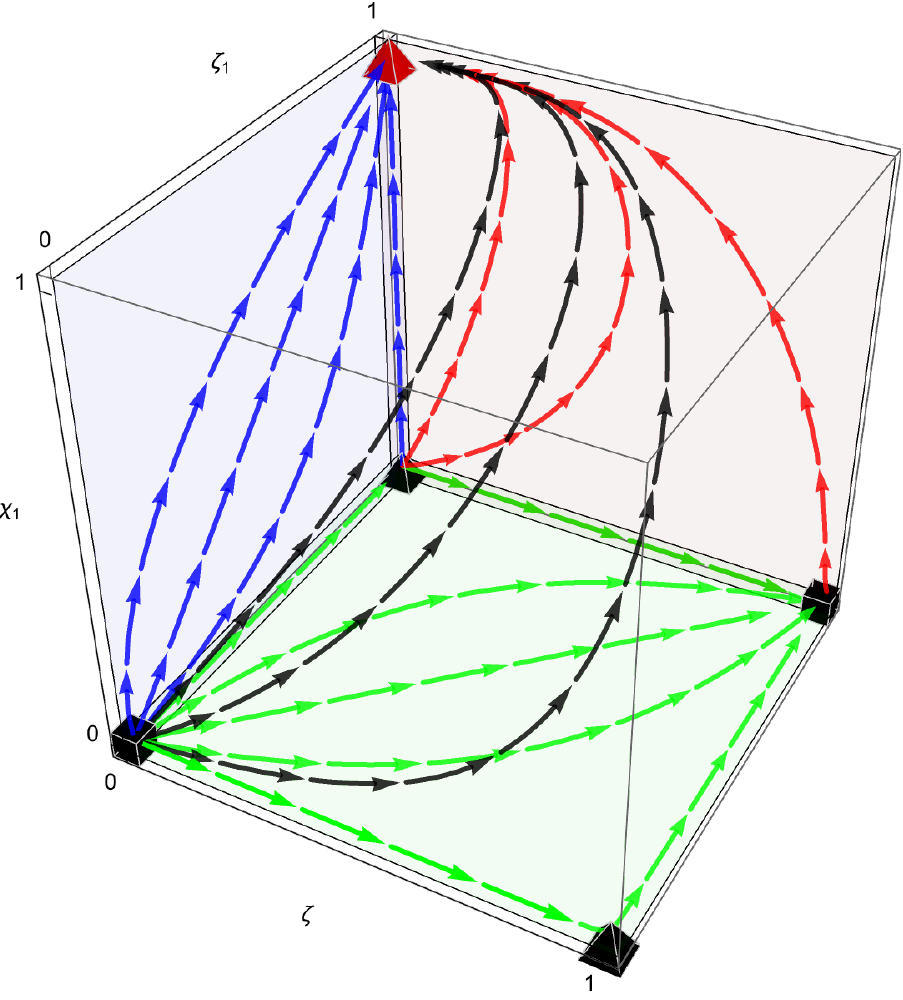}
    \label{subfig:3dflow1}}\quad
    \subfigure[]{
    \includegraphics[width=0.40\textwidth]{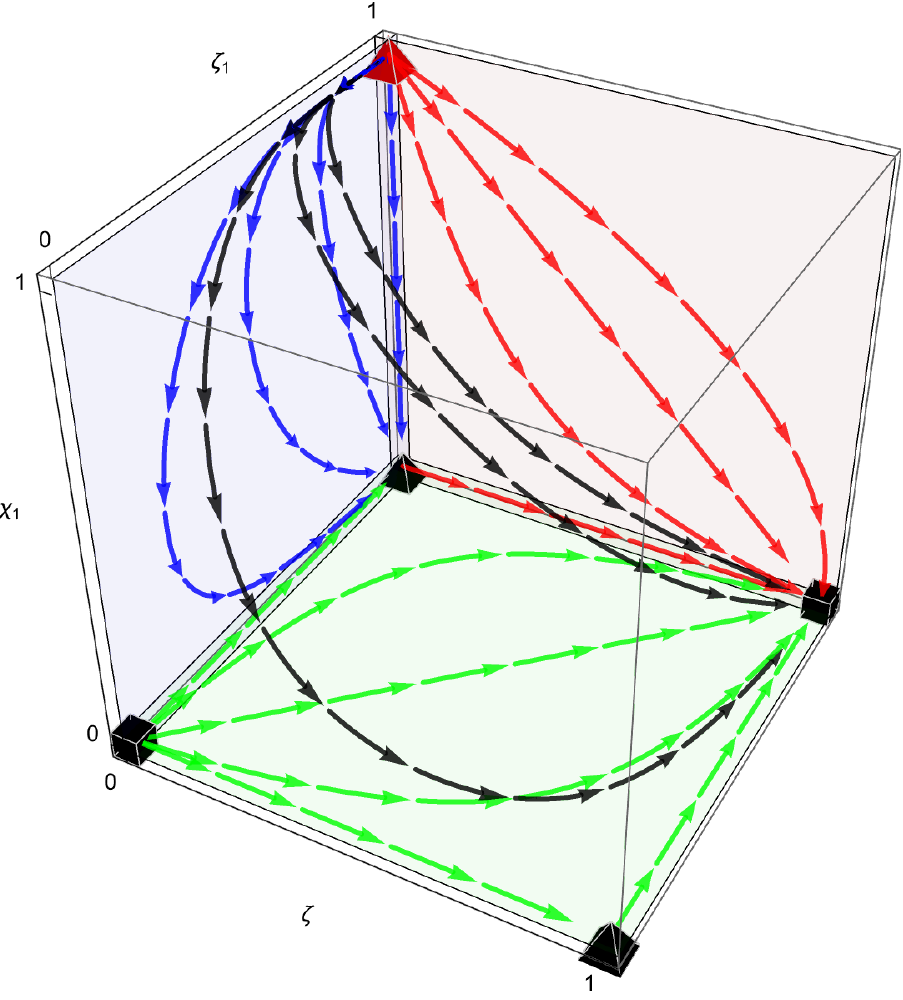}
    \label{subfig:3dflow2}} 
    \caption{\subref{subfig:3dflow1} The RG flow with $\ell=1$ in the $(\zeta_1,\zeta_k=\zeta_2=\zeta_3=\zeta_4,\chi_1)$ space. We can see that in the $\chi_1=0$ plane (no fermions), the most attractive point is the ordinary WL $W^+$. As soon as we turn on fermions, the most attractive point is the $1/2$ BPS WL with mostly positive scalar coupling matrix. \subref{subfig:3dflow2} The RG flow with $\ell=-1$ in the $(\zeta_1,\zeta_k=\zeta_2=\zeta_3=\zeta_4,\chi_1)$ space. We can see that the $1/2$ BPS Wilson loop with a mostly negative scalar coupling matrix acts as a repulsive fixed point.}
    \label{fig:3dflow}
\end{figure}

It is now interesting to compare the qualitative behavior of the RG flows in the $\ell=1$ and $\ell=-1$ cases. To this end, in figure \ref{fig:3dflow} we present a three-dimensional picture of the flows in the $(\zeta_1,\zeta,\chi_1)$ space, where $\zeta_2=\zeta_3=\zeta_4 \equiv \zeta$, for both cases. The points along the $\chi_1=0$ plane, highlighted in green in both graphs, correspond to bosonic operators. The origin is the repulsive ordinary Wilson loop $W^-$. The two neighbour points along this plane are non-BPS $SU(3)$ invariant bosonic operators with $M =\pm \text{diag}(1,-1,-1,-1)$. The fourth vertex of the green plane, diagonally opposite to the origin, is the attractive (on this plane) ordinary Wilson loop $W^+$.

Leaving the $\chi_1=0$ plane, the behavior of the flows in the two graphs start deviating. In \ref{subfig:3dflow1} all flows point towards the $1/2$ BPS Wilson loop $W^+_{1/2}$ (red triangle), confirming that this point is stable in any direction. In \ref{subfig:3dflow2} the $1/2$ BPS Wilson loop (red triangle) now corresponding to $W^-_{1/2}$ is a repulsive point, thus the most stable point is the ordinary Wilson loop $W^+$.

We can understand the different nature of the fixed points in the whole five dimensional parameter space by looking at the RG flow solutions \eqref{eq:solutionsl1} and \eqref{eq:solutionsl-1}. The most stable point will be the IR fixed point at $\mu=0$
\begin{equation}
\begin{split}
     \lim_{\mu\to 0}\begin{pmatrix}
         \zeta_1(\mu), & \zeta_k(\mu), & \chi_1(\mu)
     \end{pmatrix} =
\left\{
\begin{aligned}
    (0,1,1)\,,\qquad& \ell=+1\,. \\
    (1,1,0)\,,\qquad& \ell=-1\,.
\end{aligned}
\right.
\end{split}
\end{equation}
which corresponds to the $1/2$ BPS operator $W^+_{1/2}$ for $\ell=1$ and to the ordinary Wilson loop $W^+$ for $\ell=-1$.
Similarly, the most unstable point will be the UV fixed point at $\mu\to \infty$
\begin{equation}
\begin{split}
     \lim_{\mu\to \infty}\begin{pmatrix}
         \zeta_1(\mu), & \zeta_k(\mu), & \chi_1(\mu)
     \end{pmatrix} =
\left\{
\begin{aligned}
    (0,0,0)\,,\qquad& \ell=+1\,. \\
    (1,0,1)\,,\qquad& \ell=-1\,.
\end{aligned}
\right.
\end{split}
\end{equation}
which corresponds to the ordinary Wilson loop $W^-$ for $\ell=1$ and to the $1/2$ BPS operator $W^-_{1/2}$ for $\ell=-1$. Therefore, the nature of the fixed points enlightened in figure \ref{fig:3dflow} for three-dimensional subspaces remains unchanged in five dimensions.


\subsection{Bosonic plus two-fermion deformations}
\label{sec:doubleferm}

Now we consider deformations that involve two fermions sourced by two pairs of $(\eta, \bar{\eta})$. For concreteness, we specialize to the case where $\psi_1$ and $\psi_2$ deformations are turned on. The corresponding $\beta$-functions can be read in \eqref{eq:betachis} setting $\chi_3=\chi_4=0$, whereas the ones for the bosonic parameters are computed in detail in appendix \ref{app:ren}. They can be obtained from \eqref{eq:betasdoubleferm} setting $a=1$ and $b=2$.

As comes out from these results,  we have to distinguish between deformations with $\ell_1 =\ell_2$ or ones with $\ell_1 =-\ell_2$. In the first case, the fermionic diagram which contributes to the renormalization of the $\bar{z} M C \bar{C} z$ scalar vertex gives rise to off-diagonal terms in the scalar coupling matrix $M$. Instead, a deformation with opposite $\ell$'s does not require a non-diagonal scalar coupling matrix.
Since in what follows we restrict to the study of diagonal scalar deformations we will only consider the $\ell_1 = - \ell_2$ case.

Setting for instance $\ell_1 =1$ and $\ell_2=-1$, the $\beta$-functions are explicitly given by
\begin{equation}
\begin{split}
    \beta_{\chi_1} &= \frac{g^2N}{2\pi}\bigg( \chi_1^2-\chi_2^2 - 1 \bigg)\chi_1\,,\\
    \beta_{\chi_2}&=\frac{g^2N}{2\pi}\bigg( \chi_1^2-\chi_2^2+1\bigg)\chi_2\,,\\
    \beta_{\zeta_1} &= \frac{g^2N}{2\pi}\bigg( \zeta_1-1 + ( \chi_1^2 - \chi_2^2) \bigg)\zeta_1\,,\\
    \beta_{\zeta_2}&=\frac{g^2N}{2\pi}\left( \zeta_2-1+( \chi_1^2 - \chi_2^2)\left(1-\frac{1}{\zeta_2}\right) \right)\zeta_2\,,\\
    \beta_{\zeta_c} &= \frac{g^2N}{2\pi}\left(\zeta_c-1 +( \chi_1^2 - \chi_2^2)-\frac{\chi_1^2}{\zeta_c} \right)\zeta_c\,, \qquad c=3,4\,.
\end{split}
\end{equation}

\begin{figure}[ht]
    \centering
    \includegraphics[width=0.5\textwidth]{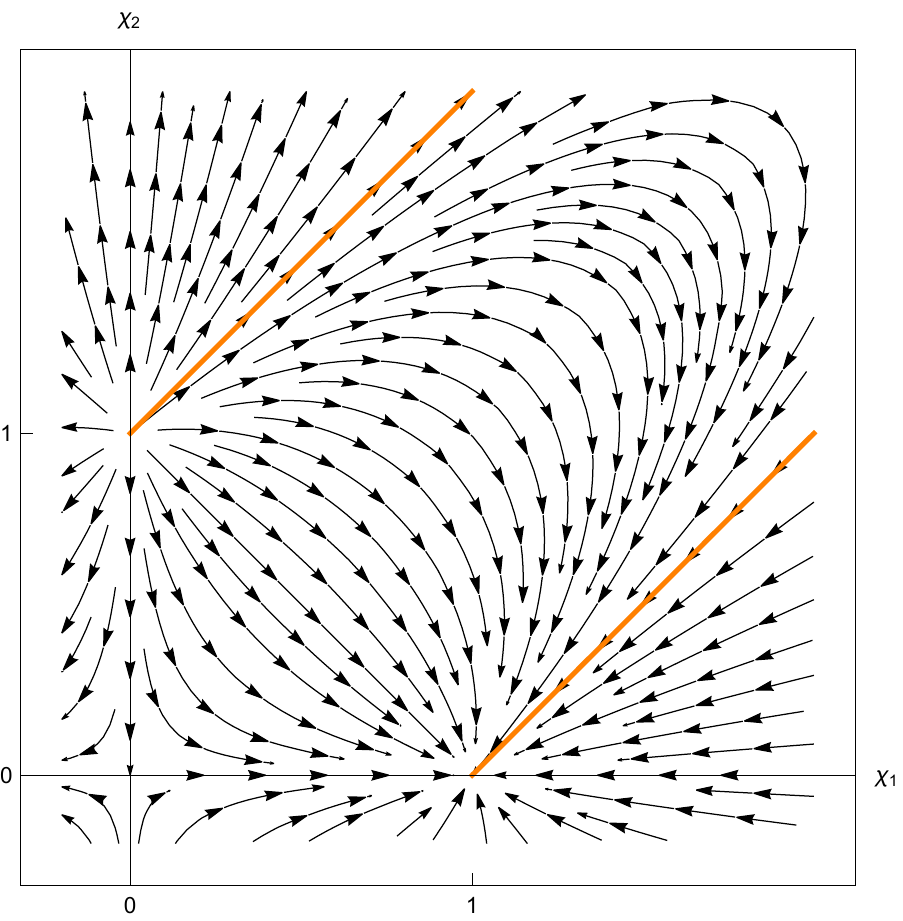}
    \caption{The RG flow in the $(\chi_1,\chi_2)$ plane for $\ell_1=+1$ and $\ell_2=-1$. The region between the two curves highlighted in orange delimits flows that start and end at fixed points $(1,0)$ and $(0,1)$.}
    \label{fig:doubleferm1}
\end{figure}

In this case the RG flow occurs in a six-dimensional space, therefore it is quite difficult to visualize it. In order to grasp some partial information, we note that at this order the fermionic $\beta$-functions do not depend on the bosonic parameters, so one can study  their flows independently of the $\zeta_i$, $i=1,\dots,4$ behavior. The fermionic flows in the $(\chi_1,\chi_2)$ plane are depicted in figure \ref{fig:doubleferm1}, where we focus only on the positive sector. As already stressed, the deformed theory is invariant under $\chi_i \to -\chi_i$, therefore the flows are simply mirrored with respect to the two axes. We highlight in orange the boundaries of the region where flows start and end at the fixed points $(1,0)$ and $(0,1)$. Since for particular choices of the $\zeta_i$ the $(1,0)$ point corresponds to $W_{1/2}^+$ while $(0,1)$ corresponds to  $W_{1/2}^-$, once more we find that  within the isolated invariant set $W_{1/2}^+$ is stable whereas $W_{1/2}^-$ is unstable.

\begin{figure}[ht]
    \centering
    \subfigure[]{
    \includegraphics[width=0.3\textwidth]{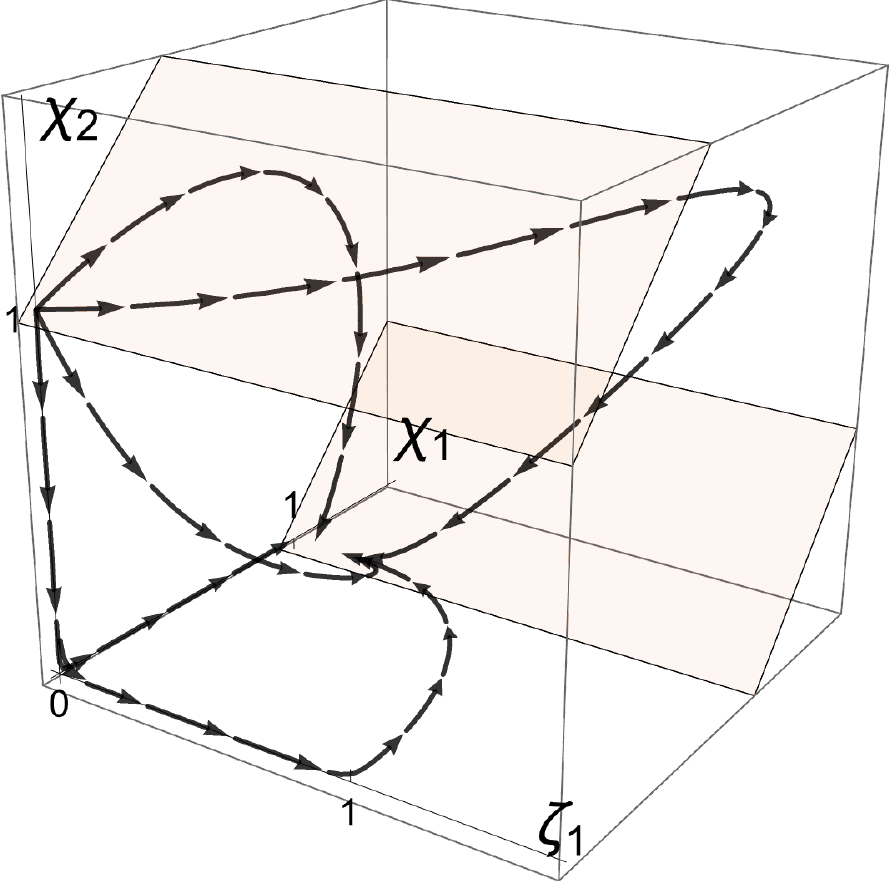}
    \label{subfig:doubleferm2-1}}
    \subfigure[]{
    \includegraphics[width=0.3\textwidth]{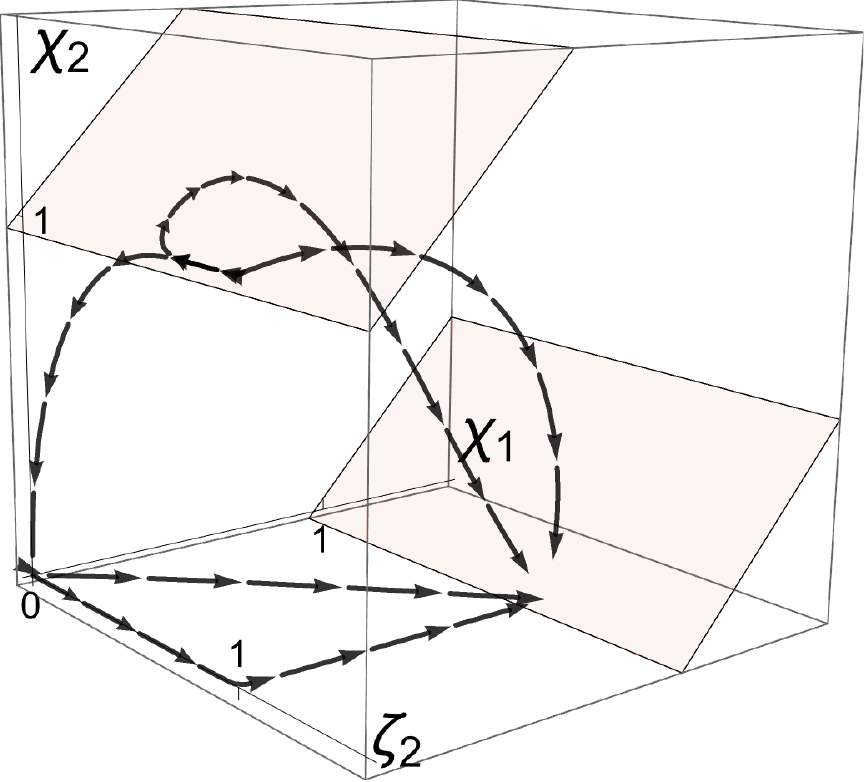}
    \label{subfig:doubleferm2-2}}
    \subfigure[]{
    \includegraphics[width=0.3\textwidth]{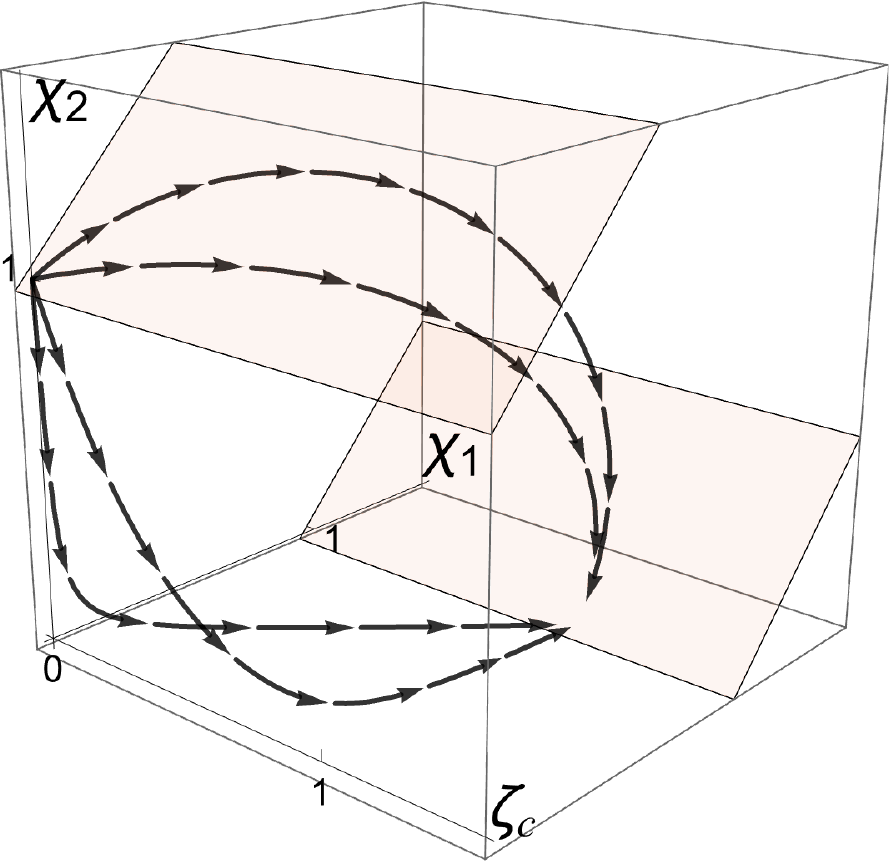}
    \label{subfig:doubleferm2-3}}
    \caption{The RG flow \subref{subfig:doubleferm2-1} in the $(\zeta_1,\chi_1,\chi_2)$ space,  \subref{subfig:doubleferm2-2} in the $(\zeta_2,\chi_1,\chi_2)$ space and \subref{subfig:doubleferm2-3} in the $(\zeta_c,\chi_1,\chi_2)$ space, $c=3,4$.}\label{fig:doubleferm2}
\end{figure}

This pattern is also very clear from figure \ref{fig:doubleferm2}, where we plot the flows on the $(\zeta_i,\chi_1,\chi_2)$ subspace for different $\zeta_i$. In orange we highlight the planes corresponding to the orange lines of figure \ref{fig:doubleferm1}. Selecting flows that lie within these two hyperplanes one easily realizes that the upper plane is repulsive whereas the lower one is attractive. 

\section{Defect theory interpretation}
\label{sec:defecttheory}

The fixed points appearing in the RG flow figures of the previous sections describe defect conformal field theories (dCFTs) living on the corresponding Wilson loops, in interaction with the ABJM bulk theory. BPS fixed points give rise to superconformal defects, whereas non-BPS points correspond to ordinary dCFTs. The RG flows have then the interesting interpretation of flows in the space of one-dimensional defect theories triggered by dynamical interactions with the bulk.  
From this perspective, they can be interpreted as being driven by marginally relevant perturbations of the defect.

In general, in the presence of deformations driven by a set of local operators $\hat{d}_i$, a defect stress tensor $T_D$ turns on, which is given by the product \cite{Cuomo:2021rkm,Giombi:2022vnz}
\begin{equation}\label{eq:stresstensor}
    T_D =\beta_i\,\hat d_i \, ,
\end{equation}
where $\beta_i$ is the $\beta$-function of the coupling associated to the $i$-th deformation. The $T_D$ operator affects the ABJM stress tensor conservation law as
\begin{equation}
    \nabla_{\mu} T^{\mu\nu} =  -\delta_D^{(2)} \left(\dot x^{\nu} \dot T_D + n_i^{\nu} D^i \right) \, ,
\end{equation}
where $\delta_D^{(2)}$ is the Dirac delta localized at the defect, $x^{\mu}$ is the embedding function describing the defect location \eqref{eq:xmu}, $n_i^{\mu}$ is a unit vector normal to the defect and $D^i$ is the displacement operator. 

From the explicit expression of $T_D$ one can compute physical quantities, as for instance the anomalous dimension of the $\hat{d}_i$ operators, and investigate the validity of a g-theorem \cite{Affleck:1991tk,Friedan:2003yc,Casini:2016fgb,Cuomo:2021rkm,Casini:2022bsu}. 

In fact, recalling that for a set of generic operators $\hat{O}_i$ the matrix of anomalous dimensions is defined as $\mu\frac{\partial \hat O_i}{\partial \mu} = -\Delta_i^{\, j}\hat O_j$, and taking into account that $T_D$ has protected mass dimension equal to one, we can apply $\mu\frac{\partial}{\partial\mu}$ to both sides of \eqref{eq:stresstensor}, obtaining
\begin{equation}
\label{eq:anomalousdimension}
    \Delta_i^{\, j} = \delta_i^{\, j} + \frac{\partial \beta^j}{\partial\zeta_i} \quad\Longrightarrow \quad \Gamma_i^{\, j} = \frac{\partial \beta^j}{\partial \zeta_i}\,,
\end{equation}
where $\Gamma_i^{\, j}$ is the anomalous dimension matrix and $\zeta_i$ is the coupling associated to $i$-th deforming operator. 

Furthermore, using the definition  \eqref{eq:stresstensor} for the defect stress tensor one can establish a one-dimensional version of the g-theorem, that is the statement that the free energy of a $\zeta$-deformed defect, ${\rm g} \equiv \log \langle W(\zeta) \rangle$, should be a monotonically decreasing function along the RG flows. In particular, it should 
satisfy ${\rm g}_{\rm UV}> {\rm g}_{\rm IR}$, where ${\rm g}_{\rm UV}$ and $ {\rm g}_{\rm IR}$ are the values of ${\rm g}$ at the UV and the IR fixed points, respectively. In \cite{Cuomo:2021rkm} this has been proven for line defects in any dimension by studying the $\zeta$-induced flow of a related observable, which is the defect entropy 
\begin{equation}
    s(\zeta)= \left( 1  + \beta_{\zeta}\frac{\partial}{\partial \zeta} \right) \log \langle W(\zeta) \rangle\,,
\end{equation}
that at the fixed points coincides with ${\rm g}$. From the relevant identity \cite{Cuomo:2021rkm}
\begin{equation}\label{eq:entropyderivative}
    \mu\frac{\partial s }{\partial \mu} = - \int d\tau_{1>2}\,\llangle T_D(\tau_1) T_D(\tau_2) \rrangle (1-\cos(\tau_1 - \tau_2))\,,
\end{equation}
which provides the mass scaling of $s$, one can conclude that the ${\rm g}$-theorem holds whenever the defect theory is reflection positive (in Euclidean signature) or unitary (in Minkowski), that is whenever $\llangle T_D(\tau_1) T_D(\tau_2) \rrangle > 0$. 

In the rest of this section we will apply eqs. \eqref{eq:anomalousdimension} and \eqref{eq:entropyderivative} to our flows in order to better investigate their nature from a defect theory perspective. 

We stress that our computations, despite being perturbative in the Chern-Simons coupling $g$, are exact in the running coupling constants $\zeta_i,\chi_i$. Therefore, from the point of view of the one-dimensional theory, the $\beta$-functions and the general behavior of the RG flows are reliable at any scale.

\subsection{Bosonic defects}

We begin by focusing on the bosonic deformations described in section \ref{sec:bosonic}. They can be interpreted as deformations of the $W^-$
bosonic dCFT\footnote{We call ``bosonic defects'' the one-dimensional theories defined on bosonic WL (no fermion fields turned on), and ``fermionic defects''
the ones defined on fermionic WLs.}
that drive the system towards IR fixed points still given by bosonic dCFTs.

In the simplest case of one-parameter deformations \eqref{eq:Mmatrix3} with $\zeta_1=\zeta_2=\zeta_3=\zeta_4\equiv \zeta$, the defect stress tensor for the deformed theory is simply given by
\begin{equation}\label{eq:TD-one}
    T_D =\beta_\zeta\,\hat d_{\zeta}= -2g^2\beta_{\zeta}C_I\bar C^I \,,
\end{equation}
where $\beta_\zeta$ is the $\beta$-function evaluated in \eqref{eq:betageneric}. 

In this case the $SU(4)$ symmetry ensures that the four operators $C_1\bar{C}^1, \cdots , C_4 \bar{C}^4$ have the same anomalous dimension. The anomalous dimension matrix in \eqref{eq:anomalousdimension} is then proportional to the identity matrix. Using \eqref{eq:betageneric} for $\beta_\zeta$, its value at $\zeta=0$ is $\Gamma = -\frac{g^2 N}{2\pi} \ \mathbb{1}$. A negative anomalous dimension signals the fact that around $W^-$ the perturbation is a weakly relevant operator that drives the defect CFT away from the fixed point. 
On the other hand, evaluating the anomalous dimension matrix for deformations around the $W^+$ fixed point in figure \ref{fig:su(4)flow}, which amounts to evaluate \eqref{eq:anomalousdimension} at $\zeta=1$, we find a positive anomalous dimension, thus the perturbation is marginally irrelevant and, consistently with the RG flow behavior, $W^+$ corresponds to an IR stable fixed point.

Moving to two-parameter deformations \eqref{eq:Mmatrix3} with $\zeta_1=\zeta_2 (\equiv \zeta_1)$ and $\zeta_3=\zeta_4 (\equiv \zeta_2)$, the preserved $SU(2) \times SU(2)$ symmetry this time insures that the $C_1\bar{C}^1$ and $C_2\bar{C}^2$ deforming operators share the same anomalous dimension $\gamma_1$,  and $C_3\bar{C}^3$ and $C_4\bar{C}^4$ share the same $\gamma_2$, as well. In principle, there might be mixing between the two pairs of operators. However, at the order we are working this is not the case. In fact, the two $\beta$-functions that can be read from \eqref{eq:betageneric} are decoupled, so that the anomalous dimension matrix \eqref{eq:anomalousdimension} is diagonal, $\Gamma = \text{diag}(\gamma_1,\gamma_1,\gamma_2,\gamma_2)$. 

Referring to figure \ref{fig:RGflow4}, we compute the values of $\gamma_1$ and $\gamma_2$ at the four fixed points. At the two black squares points, corresponding to $W^-$ (at the origin) and $W^+$, we still find that $\gamma_1(W^\pm) = \gamma_2(W^\pm) =  \pm \frac{g^2 N }{2\pi}$, in agreement with the $W^+ (W^-)$ fixed point being attractive (repulsive). 

If we consider instead the $1/6$ BPS Wilson loop associated to the scalar coupling matrix  $M = \text{diag}(-1,-1,1,1)$ and corresponding to the upper blue dot in figure \ref{fig:RGflow4}, the anomalous dimension for the $\zeta_1$ deformation is negative, $\gamma_1= - \frac{g^2 N }{2\pi}$,  whereas for the $\zeta_2$ deformation it is positive, $\gamma_2=  \frac{g^2 N}{2\pi}$. 
Therefore the $\zeta_1$ deformation of the $1/6$ BPS Wilson loop is weakly relevant, while the $\zeta_2$ one is weakly irrelevant. A similar pattern arises for the other 1/6 BPS Wilson loop associated to $M = \text{diag}(1,1,-1,-1)$ (lower blue point in figure \ref{fig:RGflow4}), simply with the signs of $\gamma_1$ and $\gamma_2$ interchanged. This is in agreement with the saddle point behavior and the directions of the flows shown in figure \ref{fig:RGflow4}. 

This analysis can be easily generalized to deformations featured by \eqref{eq:Mmatrix3} with four different parameters. In that case the is no symmetry constraining the anomalous dimension matrix. However, at the order we are working the $\beta$-functions in \eqref{eq:betageneric} are decoupled and the anomalous dimension matrix turns out to be of the form $\Gamma= {\text {diag}}(\gamma_1, \gamma_2, \gamma_3, \gamma_4)$, with four independent entries. A careful analysis reveals that $\Gamma$ is negative-definite at the $W^-$ fixed point, it is positive-definite at $W^+$, whereas at all the other fixed points it does not have a definite sign, thus confirming that they are all saddle points.  

As anticipated above, we now establish the validity of a g-theorem for the RG flows under consideration. We focus for simplicity on the one-parameter deformation that drives the RG flow between $W^-$ and $W^+$ (figure \ref{fig:su(4)flow}). Defining ${\rm g}_{\rm UV}=\log \langle W^- \rangle$ and ${\rm g}_{\rm IR}=\log \langle W^+ \rangle$, we want to check whether ${\rm g}_{\rm UV}> {\rm g}_{\rm IR}$. 

In principle, one could try a direct check by computing perturbatively ${\rm g}_{\rm UV}$ and ${\rm g}_{\rm IR}$. However, this 
would require a three-loop calculation, as up to two loops $W^\pm$ share the same expectation value. Alternatively, we can use the prescription of \cite{Cuomo:2021rkm}. If we insert the explicit expression \eqref{eq:TD-one} for $T_D$ in the scaling equation \eqref{eq:entropyderivative} for the defect entropy, at one loop we find ($\tau_{12} \equiv \tau_1 - \tau_2$)
\begin{equation}
\label{eq:defectentropy}
    \mu\frac{\partial s}{\partial \mu} = -4g^4\beta_{\zeta}^2\int d\tau_{1>2} \llangle (C_I\bar C^I)(\tau_1) (C_J\bar C^J)(\tau_2)  \rrangle(1-\cos\tau_{12})=-g^4 N^2\beta_\zeta^2 <0 \,.
\end{equation}
This means that the defect entropy decreases monotonically along the flow, leading to the expected result
\begin{equation}\label{eq:gtheorem}
   {\rm g}_{\rm UV}> {\rm g}_{\rm IR} \, .
\end{equation}

This analysis can be generalized straightforwardly to the $SU(2) \times SU(2)$ and generic bosonic flows, always leading to inequality \eqref{eq:gtheorem}. We can then conclude that within the set of bosonic deformations of the form \eqref{eq:Mmatrix3}, the ${\rm g}$-theorem is always respected.

\subsection{Fermionic defects}

We now move to the case of fermionic deformations described in section \ref{sec:fer1}. Without loss of generality, we turn on only the $\chi_1$ deformation and consider the flows connecting a $SU(3)$ invariant bosonic WL (\begin{tikzpicture}\draw[black,fill=black] (0,0) -- 
  ++(0.25,0) --
  ++(-0.125,0.25) -- cycle;
  \end{tikzpicture}) and a 1/2 BPS one (\begin{tikzpicture}\draw[red,fill=red] (0,0) -- 
  ++(0.25,0) --
  ++(-0.125,0.25) -- cycle;
  \end{tikzpicture}) (see figure \ref{fig:fermflows}). As is clear from this figure, the direction of the flow depends on the sign of $\ell$ entering the fermionic couplings \eqref{eq:etas}, thus if the ${\rm g}$-theorem is still at work,  we should expect an opposite inequality between  ${\rm g}(\begin{tikzpicture}\draw[black,fill=black] (0,0) -- 
  ++(0.25,0) --
  ++(-0.125,0.25) -- cycle;
  \end{tikzpicture})$ and ${\rm g}(\begin{tikzpicture}\draw[red,fill=red] (0,0) -- 
  ++(0.25,0) --
  ++(-0.125,0.25) -- cycle;
  \end{tikzpicture})$ in the two cases. 
Moreover, since the sign of $\ell$ also discriminates between different defects at the fixed points,\footnote{It determines the overall sign of the scalar coupling matrix at the two fixed points and the $(\eta, \bar\eta)$ couplings at the 1/2 BPS fixed point.} the two cases $\ell = \pm 1$ have to be studied separately. 

\paragraph{$\pmb{\ell=1}$ case.} In this case the $SU(3)$ invariant bosonic defect corresponds to the scalar coupling matrix
$M = {\text {diag}}(-1,1,1,1)$.
As is clear from figure \ref{fig:fermflows}, it is a UV unstable fixed point under the supermatrix deformation
\begin{equation}\label{eq:hatd}
    \hat{d} = -g \chi_1 \begin{pmatrix} 0 & \eta \bar\psi^1 \\ \psi_1 \bar\eta & 0 \end{pmatrix}\,.
\end{equation}
In fact, a straightforward application of prescription \eqref{eq:anomalousdimension} with $\beta_{\chi_1}$ as in \eqref{eq:beta1fermion} leads to a negative anomalous dimension for $\hat{d}$ at the $\chi_1 = 0$ fixed point, signaling that this is indeed a marginally relevant perturbation around the (\begin{tikzpicture}\draw[black,fill=black] (0,0) -- 
  ++(0.25,0) --
  ++(-0.125,0.25) -- cycle;
  \end{tikzpicture}) point. An analogous calculation reveals that around the BPS fixed point $W^+_{1/2}$ this is instead an irrelevant operator. 

In order to check the ${\rm g}$-theorem, we evaluate the scaling behavior of the defect entropy inserting in \eqref{eq:entropyderivative} the expression $T_D= \beta_{\chi_1} \hat{d}$, with $\hat d$ given in \eqref{eq:hatd}.\footnote{Since  coupling to fermions are defined in terms of $U(N|N)$ superconnections, the operator insertions on the defect theory are $U(N|N)$ supermatrices. Therefore, in this case $\llangle T_D(\tau_1) T_D(\tau_2) \rrangle$ in \eqref{eq:entropyderivative} includes also a trace on supermatrices.}
Evaluating the resulting expression at one loop, we find
  \begin{align}\label{eq:fermentropy}
    \mu \frac{\partial s}{\partial \mu} &= \nonumber \\
    & \hspace{-0.5cm} -g^2 \beta_{\chi_1}^2\int d\tau_{1>2} \bigg[\llangle \big( \eta \bar\psi^1 \big)(\tau_1) \big( \psi_1 \bar\eta \big)(\tau_2) \rrangle_{\ell = 1} +  \llangle \big( \psi_1 \bar\eta \big)(\tau_1) \big( \eta \bar\psi^1 \big)(\tau_2) \rrangle_{\ell =1} \bigg] (1-\cos\tau_{12}) \nonumber \\
    & = -\frac{\pi}{2} g^2 N^2 \beta_{\chi_1}^2
    \end{align}
This expression is manifestly negative, therefore the defect entropy is a monotonically decreasing function from the UV to the IR and the ${\rm g}$-theorem holds. 

\paragraph{$\pmb{\ell=-1}$ case.} From 
figure \ref{fig:fermflows} it is clear that in this case we have to consider perturbing around the 1/2 BPS operator (\begin{tikzpicture}\draw[red,fill=red] (0,0) -- 
  ++(0.25,0) --
  ++(-0.125,0.25) -- cycle;
  \end{tikzpicture})
given by a a scalar matrix $M= {\rm {diag}}(1,-1,-1,-1)$  and fermionic couplings
\begin{equation}
\label{eq:Lfchitilde}
    -i {\cal L}_F = -g\begin{pmatrix}
        0 & \eta\bar\psi^1 \\
        \psi_1\bar\eta & 0
    \end{pmatrix}\,,
\end{equation}
This corresponds to the $W^-_{1/2}$ fixed point. 
We now add the supermatrix deformation 
\begin{equation}\label{eq:hatd2}
    \hat d = g \tilde\chi_1 \begin{pmatrix}
        0 & \eta\bar\psi^1 \\
        \psi_1\bar\eta & 0
    \end{pmatrix}\,,
\end{equation}
such that the 1/2 BPS point corresponds to $\tilde\chi_1 = 0$ and the IR stable bosonic WL is obtained for $\tilde\chi_1=1$.
A simple application of prescription \eqref{eq:anomalousdimension} reveals that at $\tilde\chi_1 = 0$ this deformation is indeed marginally relevant. 

Evaluating the corresponding $\beta$-function, in this case we find 
\begin{equation}
\label{eq:betachitilde}
    \beta_{\tilde\chi_1}=-\frac{g^2 N}{2\pi}\tilde\chi_1 (\tilde\chi_1-2) (\tilde\chi_1-1)
\end{equation}
while the defect stress tensor is given by $T_D = \beta_{\tilde\chi_1} \hat d$, with $\hat d$ in \eqref{eq:hatd2}. 

Inserting $T_D$ in \eqref{eq:entropyderivative} and evaluating the expectation values at one loop, we find
\begin{align}\label{eq:fermentropy2}
    \mu \frac{\partial s}{\partial \mu} &= \nonumber \\
    & \hspace{-0.5cm} -g^2 \beta_{\tilde\chi_1}^2\int d\tau_{1>2} \bigg[\llangle \big( \eta \bar\psi^1 \big)(\tau_1) \big( \psi_1 \bar\eta \big)(\tau_2) \rrangle_{\ell = -1} +  \llangle \big( \psi_1 \bar\eta \big)(\tau_1) \big( \eta \bar\psi^1 \big)(\tau_2) \rrangle_{\ell =-1} \bigg] (1-\cos\tau_{12}) \nonumber \\
    & = \frac{\pi}{2} g^2 N^2 \beta_{\tilde\chi_1}^2
    \end{align}
In contrast with the $\ell =1$ case, this result is positive. Therefore, in the $\ell=-1$ class of deformations the ${\rm g}$-theorem is not respected. According to the general formulation of the theorem \cite{Cuomo:2021rkm}, this signals the lack of reflection positivity of this class of defects.

Summarizing, for the class of one-fermion deformations we have explicitly found that
\begin{equation}
 \mu \frac{\partial s}{\partial \mu} =\left\{
\begin{aligned}
    &\ell = +1 \quad -g^2 \beta_{\chi_1}^2 N^2 \frac{\pi}{2} \qquad \Rightarrow \quad {\rm g}_{\rm UV}(\begin{tikzpicture}\draw[black,fill=black] (0,0) -- 
  ++(0.25,0) --
  ++(-0.125,0.25) -- cycle;
  \end{tikzpicture}) > {\rm g}_{\rm IR}(\begin{tikzpicture}\draw[red,fill=red] (0,0) -- 
  ++(0.25,0) --
  ++(-0.125,0.25) -- cycle;
  \end{tikzpicture}) \, , \\
   &\ell = -1  \qquad \,g^2 \beta_{\tilde\chi_1}^2 N^2 \frac{\pi}{2} 
   \qquad \Rightarrow \quad {\rm g}_{\rm UV}(\begin{tikzpicture}\draw[red,fill=red] (0,0) -- 
  ++(0.25,0) --
  ++(-0.125,0.25) -- cycle;
  \end{tikzpicture}) < {\rm g}_{\rm IR}(\begin{tikzpicture}\draw[black,fill=black] (0,0) -- 
  ++(0.25,0) --
  ++(-0.125,0.25) -- cycle;
  \end{tikzpicture}) \,.
\end{aligned}
\right.
\label{eq:ferdefectentropy}
\end{equation}

Since this result can appear quite surprising, we provide the technical explanation of why the $\ell =1$ and $\ell =-1$ flows behave so differently. 

First, we note that possible sign differences cannot come from the $\beta$-functions since in the variation of the entropy they always appear squared. The difference comes directly from the integrated two-point functions, in particular from the different definition of the $\eta,  \bar\eta$ couplings entering $T_D$. 

In fact, looking at the first term in the integrals \eqref{eq:fermentropy}, \eqref{eq:fermentropy2} (for the second term the argument works similarly) we see that, as the result of contracting the two fermions with the propagator \eqref{eqn:propagator}, one obtains 
\begin{equation}\label{eq:TDrelation}
        \llangle T_{D}(\tau_1) T_{D}(\tau_2) \rrangle_{\ell=\pm1} \sim (\eta_1\gamma^\mu \, \bar\eta_2)_{\ell=\pm 1} \, (x_{12})_\mu
\end{equation}
where $\eta_i, \bar{\eta}_i$ stands for 
$\eta(\tau_i), \bar{\eta}(\tau_i)$ and $x_{12}^\mu \equiv (x^\mu(\tau_1) - x^\mu(\tau_2))$. 

On the other hand, from \eqref{eq:etas} it is easy to see that the $\eta, \bar\eta$ couplings satisfy the following identity
\begin{equation}\label{eq:etaid}
    (\eta_i \gamma^\mu \bar\eta_j )_\ell \,(x_{ij})_\mu= 4i \ell \sin{\frac{\tau_{ij}}{2}}.
\end{equation}
Since the rest of the factors in $\llangle T_{D}(\tau_1) T_{D}(\tau_2) \rrangle_\ell$ do not depend on $\ell$, from identity \eqref{eq:etaid} we immediately conclude that the two two-point function is proportional to $\ell$, thus it has opposite sign in the two cases
\begin{equation}
    \llangle T_{D}(\tau_1) T_{D}(\tau_2) \rrangle_{\ell=-1} = - \llangle T_{D}(\tau_1) T_{D}(\tau_2) \rrangle_{\ell=1}\,.
\end{equation}
Finally, since the results of the circle integrations are the same in two cases, we are easily led to the conclusions in \eqref{eq:ferdefectentropy}.

\section{Discussion}
\label{sec:conclusions}

We have studied defect RG flows in ABJM theory induced by bosonic and/or fermionic marginally relevant deformations. Specifically, we have focused on the subset of Wilson loops and deformations featured by diagonal scalar matrix couplings. Exploiting the one-dimensional auxiliary field description of Wilson loops to perform renormalization on the defect, we managed to compute the $\beta$-functions exactly in the deforming parameters, and at first order in the bulk ABJM coupling. Relevant features of these flows arise, which are worth recapping. 

First, starting from defects with only scalar bilinears, also referred to as bosonic defects, we have shown that generic diagonal, scalar bilinear deformations are always well-defined marginally relevant operators on the defect, as no further constraint on their structure arises along the flow. 

Regarding deformations of $SU(4)$ invariant defects, it is always possible to turn on a deformation made by a single fermion, thus necessarily breaking $SU(4)$ R-symmetry to $SU(3)$, as long as this reminiscent $SU(3)$ is also preserved by the scalar bilinears.

Progressing towards deformations containing more than one fermion, we have found in particular that turning on a two-fermion operator can be achieved only if this is accompanied by a bosonic deformation, which necessarily turns on along the flow due to the interaction of the two-fermion deformation with the bulk fields. Therefore, marginally relevant operators containing at least two fermions are necessarily given by $U(N|N)$ supermatrices with both diagonal and off-diagonal entries. Moreover, if the two fermionic deformations correspond to $(\eta, \bar{\eta})$ couplings with the same $\ell$ the bosonic deformation turns out to be non-diagonal. This is not a problem in general, as $M$ is generically non-diagonal (see for instance equation (2.10) in \cite{Castiglioni:2022yes}). However, since here we have restricted our investigation to diagonal $M$'s, we have disregarded these cases. The generalization to non-diagonal bosonic deformations should be worth investigating in the future.

Our results share some common features with similar results in $\cN=4$ SYM \cite{Polchinski_2011,Beccaria:2017rbe}, where the flow connects the ordinary loop sitting at the UV fixed point to the $1/2$ BPS fixed point in the IR. In ABJM theory the analogous pattern is realised by the mixed bosonic plus one-fermion deformation connecting $W^-$ to $W_{1/2}^+$. However, as usual, ABJM offers a much broader spectrum of fixed points. Indeed this particular realisation is only one of the many possibilities depicted in figure \ref{fig:schematics}.

One of the most striking results we have obtained is that the ordinary $W^\pm$ operators, which differ only by an apparently harmless overall sign in the scalar coupling matrix $M$, exhibit a very different nature at quantum level, being one an IR stable fixed point and the other a UV unstable one, respectively. Similarly, the $W^\pm_{1/2}$ BPS Wilson loops which differ simply by the choice of $\ell = 1$ or $-1$ in the bosonic and fermionic couplings, related to the two possible ways of solving the supersymmetry preserving constraints \cite{Cardinali:2012ru}, turn out to behave in an opposite way under deformations. In addition, the $W^-_{1/2}$ WL seems to describe a non-unitary dCFT. This is consistent with what was found in \cite{Gabai:2022vri,Gabai:2022mya} for mesonic line operators in Chern-Simons-matter theories. While we have provided a perturbative explanation of these features, it would be nice to have a more physical interpretation.

Focusing on 1/2 BPS operators, we recall that in field theory $W_{1/2}^+$ and $W_{1/2}^-$ can be obtained by a Higgsing construction that makes use of heavy $W$-particles and $W$-antiparticles, respectively \cite{Lietti:2017gtc} ($W$ and $\tilde W$ there). This provides already some indication that they could give rise to physically different dCFTs. 
This seems to be confirmed also holographically. In fact, as shown in \cite{Lietti:2017gtc}, the two $1/2$ BPS fixed points have different classical M-theory duals:\footnote{A first quantum perturbative calculation  around this classical configuration can be found in \cite{Giombi:2023vzu}.} while $W_{1/2}^+$ is dual to a M2-brane configuration in ${\rm AdS}_4 \times {\rm S}^7/{\mathbb Z}_k$, $W_{1/2}^-$ is described by an anti-M2-brane. Although this seems to be consistent with our findings, it is not clear how to exactly relate the opposite behavior under RG flow found in this paper to the two different holographic descriptions. This  is a very interesting question that deserves further investigation. 

A less clear picture exists for ordinary WLs. 
 As is well-known, the holographic descriptions of different Wilson loops with different amount of supersymmetry differ by the set of boundary conditions of the corresponding string configuration in the internal space being Dirichlet or Neumann. A naive comparison with the $\cN=4$ SYM setting suggests that our ``ordinary'' $SU(4)$ Wilson loop operators should be described by strings satisfying pure Neumann boundary conditions, that is strings smeared over $\mathbb{CP}^3$. It is not clear, however, how a distinction between $W^+$ and $W^-$ should arise from the dual point of view. Of course, a holographic input would allow for a better understanding of our results.

\begin{figure}[ht]
    \centering
    \includegraphics[width=.4\textwidth]{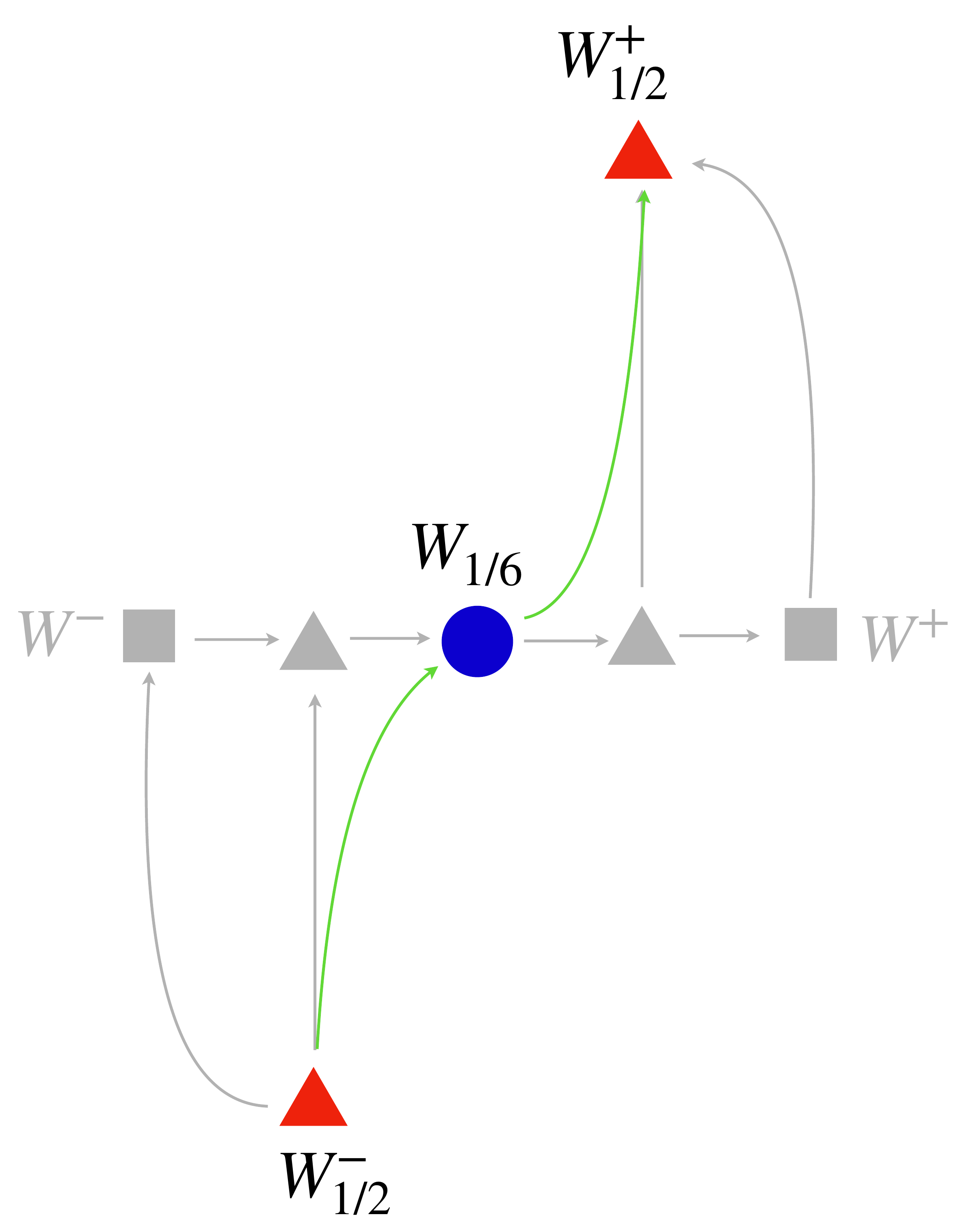}
    \label{subfig:schematics-enriched-a}
    \caption{Representation of enriched flows connecting $W_{1/6}$ and $W_{1/2}^\pm$.}
    \label{fig:schematics-enriched}
\end{figure}

Our current discussion includes the enriched flows studied in \cite{Castiglioni:2022yes}, when we focus on the particular case of BPS RG trajectories connecting $W_{1/6}$ and $W_{1/2}^\pm$. These are depicted as green lines in figure \ref{fig:schematics-enriched}. The green curve connecting $W_{1/6}$ and $W^+_{1/2}$ corresponds to the parabolic RG trajectory in figure \ref{subfig:1fermflow3} given by $\zeta = \chi_1^2$. It consists of a smooth interpolation made by a continuum of $1/6$ BPS fermionic loops, thus the name ``enriched flow''. It corresponds exactly to the green trajectory in figure 13 of \cite{Castiglioni:2022yes}. The enriched trajectory connecting $W^-_{1/2}$ and $W_{1/6}$, instead, is not directly included in the analysis of section \ref{sec:generic}, as the $W_{1/6}$ fixed point should correspond to the one considered there, but with $M \to -M$. Of course, it is straightforward to duplicate the analysis of section \ref{sec:generic} in this case and pinpoint the enriched trajectory. The $W^-_{1/2}$ operator, which turns out to be a repulsive fixed point, coincides with ${\cal W}^{II}_{1/2}$ of \cite{Castiglioni:2022yes} upon an R-symmetry rotation. We thus complete the analysis of \cite{Castiglioni:2022yes} with new information about enriched RG flows connecting the repulsive ${\cal W}^{\rm II}_{1/2}$. For the interested reader, we discuss details in appendix \ref{app:roadmap}.

Our results are perturbative in the ABJM coupling, so higher order corrections to the $\beta$-functions might potentially modify the spectrum of fixed points. Checking what happens to them at higher loops  is an interesting open question that would be worth addressing.

\vskip 40pt 

\section*{Acknowledgements}

We are grateful to Diego Correa and Guillermo Silva for discussions. LC, SP and MT are partially supported by the INFN grant {\it Gauge Theories, Strings and Supergravity (GSS)}.
DT is supported in part by the INFN grant {\it Gauge and String Theory (GAST)}. DT would like to thank FAPESP’s partial support through the grants 2016/01343-7 and 2019/21281-4.

\newpage
\appendix

\section{Conventions and Feynman rules}
\label{app:abjm}

For ABJM theory we follow the conventions in \cite{Bianchi:2014laa}. We work in three-dimensional Euclidean space with coordinates $x^{\mu}=(x^0,x^1,x^2)$. The three-dimensional gamma matrices are defined as
\begin{equation}\label{eq:gamma}
    (\gamma^{\mu})^{ \ \beta}_{ \alpha}=(-\sigma^3,\sigma^1,\sigma^2)_\alpha^{\ \beta}\,,
\end{equation}
with $(\sigma^{i} )^{ \ \beta}_{  \alpha}$ ($\alpha,\beta=1,2$) being the Pauli matrices, such that $\gamma^{\mu}\gamma^{\nu}=\delta^{\mu\nu}+i\epsilon^{\mu\nu\rho}\gamma_{\rho}$, where $\epsilon^{123}=\epsilon_{123}=1$ is totally antisymmetric. Spinorial indices are lowered and raised as $(\gamma^{\mu})^{\alpha}_{\ \beta}=\epsilon^{\alpha\gamma}(\gamma^{\mu})^{\ \delta}_{\gamma} \epsilon_{\beta\delta}$, with $\epsilon_{12}=-\epsilon^{12}=1$. The Euclidean action of $U(N)_k\times U(N)_{-k}$ ABJM theory is
\begin{equation}
\label{eq:ABJMaction}
\begin{split}
    S_{\textrm{ABJM}}=&\frac{k}{4\pi} \int d^3 x\,  \epsilon^{\mu\nu\rho}\Big\{ -i\text{Tr}\left( A_{\mu}\partial_{\nu}A_{\rho} +\frac{2i}{3}A_{\mu}A_{\nu}A_{\rho} \right)+i\text{Tr}\left( \hat A_{\mu}\partial_{\nu}\hat A_{\rho} +\frac{2i}{3}\hat A_{\mu}\hat A_{\nu}\hat A_{\rho} \right) \\ & + \text{Tr}\left[ \frac{1}{\xi}(\partial_{\mu}A^{\mu})^2 - \frac{1}{\xi}(\partial_{\mu}\hat A^{\mu})^2 +\partial_{\mu} \bar c D^{\mu}c-\partial_{\mu} \bar{\hat c}D^{\mu}\hat c\right] \Big\} \\ & + \int d^3x \text{Tr}\left[ D_{\mu} C_I D^{\mu} \bar C^I +i\bar\psi^I \gamma^{\mu}D_{\mu} \psi_I \right]\\ & \begin{split}- \frac{2\pi i}{k}\int d^3 x \text{Tr}\Big[& \bar C^I C_I \psi_J \bar\psi^J - C_I \bar C^I \bar\psi^J \psi_J + 2C_I\bar C^J \bar\psi^I\psi_J \\ &-2\bar C^I C_J \psi_I \bar\psi^J - \epsilon_{IJKL}\bar C^I \bar \psi^J \bar C^K \bar \psi^L +\epsilon^{IJKL} C_I \psi_J C_K \psi_L \Big]+ S^{\text{bos}}_{\text{int}}\,,\end{split}
\end{split}
\end{equation}
with covariant derivatives defined as
\begin{alignat*}{3}
    &D_{\mu} C_I &&= \partial_{\mu} C_I +i A_{\mu} C_I -i C_I \hat A_{\mu}\,, \qquad D_{\mu} \bar C^I &&=\partial_{\mu} \bar C^I -i \bar C^I A_{\mu} + i\hat A_{\mu} \bar C^I\,, \\
    &D_{\mu}\bar\psi^I &&=\partial_{\mu} \bar\psi^I + iA_{\mu} \bar\psi^I - i\bar\psi^I \hat A_{\mu}\,, \qquad D_{\mu} \psi_I &&=\partial_{\mu}\psi_I -i\psi_I A_{\mu} +i\hat A_{\mu} \psi_I\,.
\end{alignat*}
We work in Landau gauge for vector fields and in dimensional regularization with $d=3-2\epsilon$. The tree-level propagators are (with $g=\sqrt{2\pi/k}$)
\begin{equation}
\label{eqn:propagator}
    \begin{split}
        \langle (A_{\mu})_p^{\ q}(x)(A_{\nu})_r^{\ s}(y)\rangle^{(0)} &=\delta_p^s\delta_r^q\, i g^2 \, \frac{\Gamma(\frac{3}{2}-\epsilon)}{2\pi^{\frac{3}{2}-\epsilon}}\frac{\epsilon_{\mu\nu\rho}(x-y)^{\rho}}{|x-y|^{3-2\epsilon}},\\
        \langle (\hat A_{\mu})_{\hat p}^{\ \hat q}(x)(\hat A_{\nu})_{\hat r}^{\ \hat s}(y)\rangle^{(0)} &=-\delta_{\hat p}^{\hat s}\delta_{\hat r}^{\hat q} \, i g^2 \, \frac{\Gamma(\frac{3}{2}-\epsilon)}{2\pi^{\frac{3}{2}-\epsilon}}\frac{\epsilon_{\mu\nu\rho}(x-y)^{\rho}}{|x-y|^{3-2\epsilon}},\\
        \langle  (\psi_I^{\alpha})_{\hat i}^j(x) (\bar\psi_{\beta}^J)_k^{\hat l} (y) \rangle^{(0)} & = -i\delta_I^J\delta_i^{\hat l}\delta_k^{j } \frac{\Gamma(\frac{3}{2}-\epsilon)}{2\pi^{\frac{3}{2}-\epsilon}}\frac{(\gamma_{\mu})^{\alpha}_{\ \beta}(x-y)^{\mu}}{|x-y|^{3-2\epsilon}}\\ & =i\delta_I^J\delta_i^{\hat l}\delta_k^{j } (\gamma_{\mu})^{\alpha}_{\ \beta}\partial_{\mu}\left( \frac{\Gamma(\frac{1}{2}-\epsilon)}{4\pi^{\frac{3}{2}-\epsilon}} \frac{1}{|x-y|^{1-2\epsilon}}\right), \\
        \langle (C_I)_i^{\hat j}(x)(\bar C^J)_{\hat k}^l(y)\rangle^{(0)} &= \delta_I^J \delta_i^l \delta_{\hat k}^{\hat j}  \frac{\Gamma(\frac{1}{2}-\epsilon)}{4\pi^{\frac{3}{2}-\epsilon}}\frac{1}{|x-y|^{1-2\epsilon}} ,
    \end{split}
\end{equation}
while the one-loop propagators read
\begin{equation}
\label{eqn:onelooppropagator}
\begin{split}
    \langle (A_{\mu})_p^{\ q}(x) (A_{\nu})_r^{s}(y) \rangle^{(1)} &= \delta_p^s\delta_r^q \, g^4 N \, \frac{\Gamma^2(\frac{1}{2}-\epsilon)}{4\pi^{3-2\epsilon}}\left[ \frac{\delta_{\mu\nu}}{|x-y|^{2-4\epsilon}}-\partial_{\mu}\partial_{\nu}\frac{|x-y|^{2\epsilon}}{4\epsilon(1+2\epsilon)} \right] ,\\
    \langle (\hat A_{\mu})_{\hat p}^{\ \hat q}(x) (\hat A_{\nu})_{\hat r}^{ \hat s}(y) \rangle^{(1)} &= \delta_{\hat p}^{\hat s}\delta_{\hat r}^{\hat q} \, g^4 N \, \frac{\Gamma^2(\frac{1}{2}-\epsilon)}{4\pi^{3-2\epsilon}}\left[ \frac{\delta_{\mu\nu}}{|x-y|^{2-4\epsilon}}-\partial_{\mu}\partial_{\nu}\frac{|x-y|^{2\epsilon}}{4\epsilon(1+2\epsilon)} \right] \, .
\end{split}
\end{equation}
The latin indices are color indices. For instance, $(A_{\mu})_p^{\ q} \equiv A_\mu^a (T^a)_p^{\ q}$ where $T^a$ are $U(N)$ generators in the fundamental representation.


\section{Wilson loops in ABJM theory}
\label{app:BPSWL}

\paragraph{BPS Wilson loops.} In ABJM theory there is a wide set of BPS operators one can construct, which may carry some parametric dependence that allows for a continuous interpolation between different observables. Such set was the main character of \cite{Castiglioni:2022yes}. Here we focus instead on non-BPS flows and only a few BPS operators within the aforementioned set appear in our discussion. We list these below, together with the pictorial representation used in the figures throughout the paper.

We start from the simplest realization where the loops are charged under a single node and are bosonic. In this case we may have operators separately charged under the first node, associated with $A$, or the second node, associated with $\hat{A}$, of the ABJM quiver. In Euclidean signature they are explicitly given by
\begin{equation}
\begin{tikzpicture}
  \filldraw [blue] (0,0) circle (3pt);
  \end{tikzpicture}\,\,\,\,\left\{
\begin{aligned}
    W_{1/6} &= \Tr\cP\exp\bigg(-i \oint d\tau\, \cA \bigg)\,, \quad \cA \equiv  A_{\mu} \dot x^{\mu} - ig^2\vert\dot{x}\vert {M_I}^J C^I \bar C_J\,, \\
   \hat W_{1/6} &= \Tr\cP\exp\bigg(-i \oint d\tau\, \hat{\cA} \bigg)\,, \quad \hat{\cA}\equiv \hat{A}_{\mu} \dot x^{\mu} - ig^2\vert\dot{x}\vert {M_I}^J \bar C_J C^I ,
\end{aligned}
\right.
\label{eq:WLs}
\end{equation}
where $g^2=2\pi/k$. In both cases for $M=\pm\diag(-1,-1,1,1)$ (plus any other permutation of the diagonal entries) these operators become $1/6$ BPS. Whenever present in the figures, these $SU(2)\times SU(2)$ symmetric objects are represented as blue dots/balls.

In addition, we also have a pair of $1/2$ BPS operators that are charged under both nodes of the quiver and are therefore  written in terms of a superconnection $\cL$. Explicitly, we have
\begin{equation}
\label{eq:W1/2pm}
\begin{tikzpicture}
  \draw[red,fill=red] (0,0) -- 
  ++(0.25,0) --
  ++(-0.125,0.25) -- cycle;
  \end{tikzpicture}\,\,\,\,\;
\begin{aligned}
    W^\pm_{1/2} &=  \Tr\cP\exp\bigg(-i \oint d\tau\, \cL^\pm \bigg)\,, \quad \cL^\pm \equiv  \begin{pmatrix} \cA && -ig\eta\bar\psi^1 \\
    -ig\psi_1\bar\eta && \hat{\cA}
    \end{pmatrix}\,,
\end{aligned}
\end{equation}
where $\cA$ and $\hat\cA$ are as in \eqref{eq:WLs} but with scalar coupling matrix $M=\ell\diag(-1,1,1,1)$. Correspondingly the commuting spinors $\eta$ and $\bar\eta$ are those in \eqref{eq:etas}. $W^+_{1/2}$ is then obtained by setting $\ell= 1$, while $W^-_{1/2}$ corresponds to $\ell=-1$. In our figures these $SU(3)$ symmetric BPS objects are represented as red triangles/pyramids.

\paragraph{Non-BPS Wilson loops.}

We find fixed points corresponding to non-BPS operators that are $SU(3)$ invariant. As in the BPS case, for bosonic operators these can be defined separately for each node of the ABJM quiver,
\begin{equation}
\begin{tikzpicture}
  \draw[black,fill=black] (0,0) -- 
  ++(0.25,0) --
  ++(-0.125,0.25) -- cycle;
  \end{tikzpicture}\,\,\,\,\left\{
\begin{aligned}
    W &= \Tr\cP\exp\bigg(-i \oint d\tau\, \cA \bigg)\,, \quad \cA =  A_{\mu} \dot x^{\mu} - ig^2\vert\dot{x}\vert {M_I}^J C^I \bar C_J\,, \\
   \hat W &= \Tr\cP\exp\bigg(-i \oint d\tau\, \hat{\cA} \bigg)\,, \quad \hat{\cA}= \hat{A}_{\mu} \dot x^{\mu} - ig^2\vert\dot{x}\vert {M_I}^J \bar C_J C^I .
\end{aligned}
\right.
\label{eq:bosonicsu3}
\end{equation}
with $M=\pm\diag(-1,1,1,1)$ or permutations. Such options should be equivalent up to R-symmetry rotations.

In addition, we also find $SU(3)$ fixed points that are fermionic, in which case they are defined as
\begin{equation}
    \label{eq:unusualsu3}
    \begin{tikzpicture}
  \draw[black,fill=black] (0,0) -- 
  ++(0.25,0) --
  ++(-0.125,0.25) -- cycle;
  \end{tikzpicture}\,\,\,\,\;
\begin{aligned}
    W &=  \Tr\cP\exp\bigg(-i \oint d\tau\, \cL \bigg)\,, \quad \cL \equiv  \begin{pmatrix} \cA && -ig\eta\bar\psi^1 \\
    -ig\psi_1\bar\eta && \hat{\cA}
    \end{pmatrix}\,,
\end{aligned}
\end{equation}
with $\cal A, \hat{\cal A}$ given as in \eqref{eq:WLs} but now with $M=\ell\diag(-1,-3,-3,-3)$, and $\eta$, $\bar\eta$ given in \eqref{eq:etas}. Such points are shown in figures \ref{subfig:1fermflow1} and \ref{subfig:2fermflow2}.

Finally, there are also $SU(2)$ fermionic fixed points, denoted as black circles. These are defined as
\begin{equation}
    \label{eq:unusualsu2}
    \begin{tikzpicture}
  \filldraw [black] (0,0) circle (3pt);
  \end{tikzpicture}\,\,\,\,\;
\begin{aligned}
    W &=  \Tr\cP\exp\bigg(-i \oint d\tau\, \cL \bigg)\,, \quad \cL \equiv  \begin{pmatrix} \cA && -ig\eta\bar\psi^1 \\
    -ig\psi_1\bar\eta && \hat{\cA}
    \end{pmatrix}\,,
\end{aligned}
\end{equation}
with $M=\diag(-1,1,-3,-3)$ or $M=\diag(-1,-3,1,1)$ in $\cal A, \hat{\cal A}$. They appear in figures \ref{subfig:1fermflow2} and \ref{subfig:1fermflow3}.


\section{Renormalization computations}
\label{app:ren}
In this section we report the explicit calculation of the $\beta$-functions. We follow what was done for the $1/24$ BPS interpolating Wilson loop in \cite{Castiglioni:2022yes}.
We study the renormalization of the quantum field theory defined by the following effective action
\begin{equation}
\hspace{-0.5cm}  
\label{eqn:effS}
\begin{split}
   S_\textrm{eff} = S_\textrm{ABJM} +\int d\tau  \bar\Psi \left( 
\partial_{\tau} + i\cL \right)\Psi \,,
\end{split}
\end{equation}
where $\cL$ is the deformed Wilson loop (super)connection and we have defined the one-dimensional Grassmann odd superfield
\begin{equation}\label{eq:oddmatrix}
    \Psi = \begin{pmatrix} z & \varphi \\ \tilde \varphi& \tilde z\end{pmatrix}\,, \qquad \bar\Psi = \begin{pmatrix} \bar z & \bar{\tilde\varphi} \\ \bar{\varphi}& \bar{\tilde z}\end{pmatrix}\,,
\end{equation}
where $z$ ($\tilde z$) and $\varphi$ ($\tilde \varphi$) are a spinor and a scalar, respectively, in the fundamental representation of $U(N)$. 


The tree-level propagators of the one-dimensional fields are
\begin{equation}
\begin{alignedat}{3}
    & \includegraphics[width=0.2\textwidth]{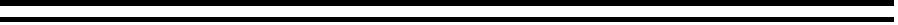} &&= \langle z^i(\tau_1)\bar z_j(\tau_2) \rangle &&= \delta_j^i \, \theta(\tau_1-\tau_2)\,, \\
    & \includegraphics[width=0.2\textwidth]{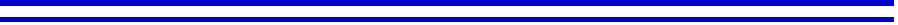} &&= \langle {\tilde z}^{\hat i}(\tau_1)\bar{\tilde z}_{\hat j}(\tau_2) \rangle &&= \delta_{\hat j}^{\hat i} \,  \theta(\tau_1-\tau_2)\,, \\
    & \includegraphics[width=0.2\textwidth]{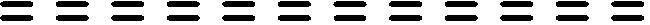} &&= \langle \varphi^i(\tau_1)\bar{\varphi}_j(\tau_2) \rangle &&= \delta_j^i \,  \theta(\tau_1-\tau_2)\,, \\
    & \includegraphics[width=0.2\textwidth]{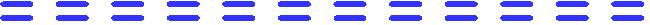} &&= \langle \tilde \varphi^{\hat i}(\tau_1)\bar{\tilde \varphi}_{\hat j}(\tau_2) \rangle &&= \delta_{\hat j}^{\hat i} \,  \theta(\tau_1-\tau_2)\,.
\end{alignedat}
\end{equation}
In order to renormalize the theory, for each one-dimensional field $\phi=\{\varphi,\tilde{\varphi},z,\tilde{z}\}$ we introduce the corresponding renormalization function as $\phi=Z_\phi^{-\frac{1}{2}}\phi_0$,
where $\phi_0$ stands for the bare quantity.

Focusing on the one-loop renormalization of the $\zeta_i$ parameters in the bosonic deformation \eqref{eq:Mmatrix3}, we need to consider the scalar vertex $\bar z C \bar C z$ (similarly for the other one-dimensional fields). We define the renormalization functions $Z_{\zeta_i}$ such that
\begin{equation}\label{eq:renF}
    (\zeta_i)_0 = Z_{\zeta_i} \zeta_i=(1+\delta_{\zeta_i}) \zeta_i\,.
\end{equation}
We write the action \eqref{eqn:effS} as a function of the renormalized parameters adding the counterterm $g^2 \delta M_I^{\ J} \bar{z} C_J \bar{C}^I z$ with
\begin{equation}
    \delta M_I^{ \ J} = 2 \begin{pmatrix} \delta_{\zeta_1}\zeta_1 & 0 & 0 & 0 \\ 0 & \delta_{\zeta_2}\zeta_2 & 0 & 0 \\ 0 & 0 & \delta_{\zeta_3}\zeta_3 & 0 \\ 0 & 0 & 0 & \delta_{\zeta_4}\zeta_4 \end{pmatrix}\,.
\end{equation}
For the other one-dimensional scalar field vertices the calculations are similar.

\subsection{Bosonic deformation}\label{sec:pert1}

\begin{figure}[t]
    \centering
    \subfigure[]{
    \includegraphics[width=0.25\textwidth]{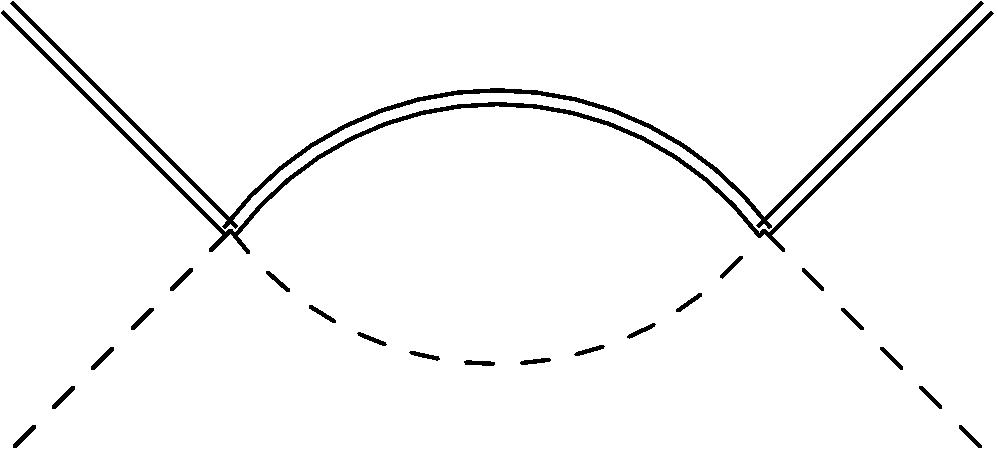}
    \label{subfig:diaga}} \qquad\qquad
    \subfigure[]{
    \includegraphics[width=0.18\textwidth]{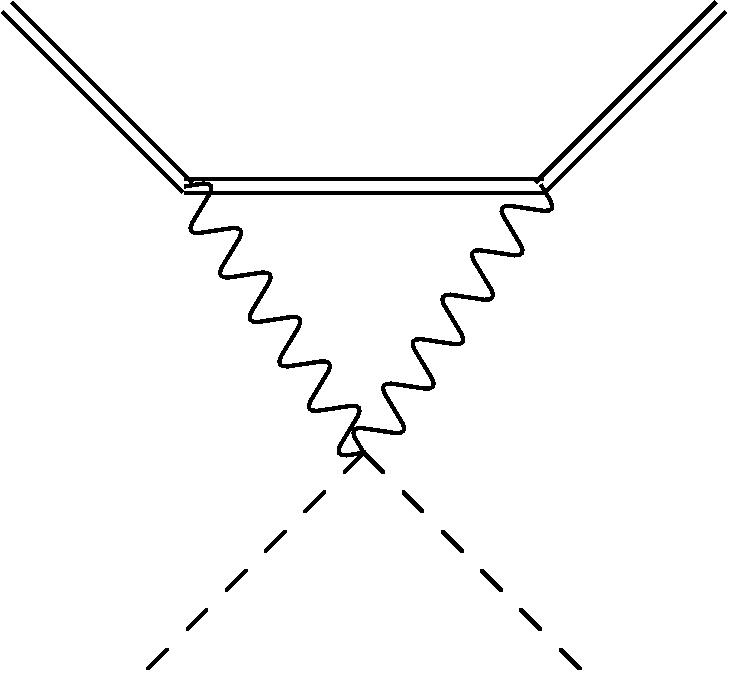}
    \label{subfig:diagb}}  \qquad\qquad
    \subfigure[]{
    \includegraphics[width=0.18\textwidth]{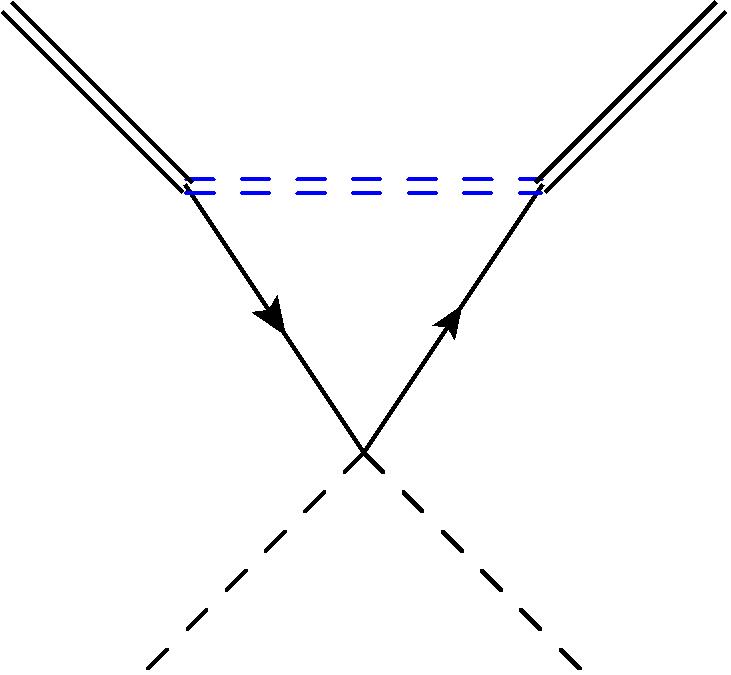}
    \label{subfig:diagc}} 
    \caption{One-loop corrections to the $\bar z C \bar C z$ vertex. \subref{subfig:diaga} and \subref{subfig:diagb} are purely bosonic diagrams, while \subref{subfig:diagc} contains fermions.}
    \label{fig:diag1}
\end{figure}

The divergent scalar one-loop diagrams related to the $\bar{z}C\bar{C}z$ vertex are depicted in figures \ref{subfig:diaga} and \ref{subfig:diagb}. We refer to \cite{Castiglioni:2022yes} for the explicit computation. We find
\begin{equation}
\begin{split}
     & \Gamma^{\rm \ref{subfig:diaga}} =  - \frac{g^4N}{8\pi\epsilon} M_{I}^{ \ K} M_{K}^{\ J} \int d\tau\,  \bar z \, C_J \bar C^I  z\,, \\
    &\Gamma^{\rm \ref{subfig:diagb}} =   \frac{g^4N}{8\pi\epsilon} \int d\tau\,  \bar z \, C_I \bar C^I z \,,
\end{split}
\end{equation}
where $M$ is the deformed scalar matrix. 

These are the only divergent contributions to the vertex, as in the absence of fermions there is no field function renormalization at one loop. 
In minimal subtraction scheme the counterterm $\delta M_I^{\ J}$ is then obtained by imposing
\begin{equation}
    0= g^2\left[ \delta M_I^{\ J} -\frac{g^2N}{8\pi\epsilon}M_{I}^{ \ K} M_{K}^{\ J} + \frac{g^2N}{8\pi\epsilon} \delta_I^{J} \right]\int d\tau \bar z C_J \bar C^I z\,.
    \label{eqn:count=0}
\end{equation}
From each diagonal element of \eqref{eqn:count=0} we find the corresponding counterterm
\begin{equation}\label{eqn:z1}
    \delta_{\zeta_i}\zeta_i = \frac{g^2 N}{4\pi\epsilon}(\zeta_i-1)\zeta_i\,.
\end{equation}

In dimensional regularization, $d=3-2\epsilon$ with minimal subtraction, the $\zeta_i$ parameters are dimensionless while $g^2$ has dimension $\Delta_{g^2}=2\epsilon$. We can then write the $\beta$-functions for $\zeta_i$ as
\begin{equation}
    \beta_{\zeta_i} = 2g^2 \frac{\partial K_{\zeta_i}}{\partial g^2}\,,
\end{equation}
where $K_{\zeta_i}$ are the coefficients of the divergent part of $(\zeta_i)_0$ as a function of $\zeta_i$ (see for instance equation (3.43) in \cite{Castiglioni:2022yes}). Therefore, combining \eqref{eq:renF} with \eqref{eqn:z1}, we eventually find
\begin{equation}
     \beta_{\zeta_i} = \frac{g^2N}{2\pi}(\zeta_i-1)\zeta_i\,.
\end{equation}

\subsection{Fermionic deformation}
\label{app:fermdef}

The presence of fermions in the Wilson loop superconnection gives rise to one extra divergent diagram, figure \ref{subfig:diagc}, which contributes to the $\bar z C \bar C z$ vertex renormalization. Moreover we also have to take into consideration the renormalization of the $z$ field.

Diagram \ref{subfig:diagc} gives two different contributions. The first one arises from the ABJM vertex $C_I\bar C^I \bar\psi^J\psi_J$. This contribution is not affected by the two possible values of $\ell$ in the definition of $\eta,\bar\eta$. Turning on only  the $i$th deformation we find
\begin{equation}
    \Gamma^{\rm \ref{subfig:diagc}}_1 = \frac{g^2N}{4\pi\epsilon}  \chi_i^2 \int d\tau \, \bar z C_I \bar C^I z\, , \qquad i =1,2,3,4 \, .
\end{equation}
This is a divergent contribution to the diagonal scalar coupling matrix $M_I^{\; J}$ to be included in \eqref{eqn:count=0}.
More generally, if we turn on more than one $\chi$-deformation the overall contribution will be the sum of $\Gamma^{\rm \ref{subfig:diagc}}_1$, one for each $\chi_i$.

The second contribution comes from using the ABJM vertex $C_I\bar C^J \bar\psi^I \psi_J$. Also in this case, if we turn on only one fermion the result is not affected by the choice of   $\ell$. For the $i$th deformation we find 
\begin{equation}
\label{eq:figc1}
   \Gamma^{\rm \ref{subfig:diagc} }_2 = -\frac{g^2 N}{2\pi\epsilon} \chi_i^2 \int d\tau \, \bar z C_I \bar C^I z\, , \qquad i =1,2,3,4 \, . 
\end{equation}

However, from the point of view of the $\ell$ dependence this time things drastically change if we turn on more than one fermion.  In fact, if we consider a two-fermion deformation corresponding to $\chi_a$ with $\ell_a$ and $\chi_b$ with $\ell_b$ (with $a\ne b$) we find
\begin{equation}
\label{eq:figc2}
\Gamma^{\rm \ref{subfig:diagc} }_2 = -\frac{g^2 N}{2\pi\epsilon}\bigg( \chi_a^2 \, \delta_I^a \delta_a^J +  \chi_b^2 \, \delta_I^b \delta_b^J - \frac{\vert \ell_a+\ell_b\vert}{2}\chi_a\chi_b \, \big(\delta_I^a \delta_b^J +\delta_I^b \delta_a^J\big)\bigg)\int d\tau \, \bar z C_J \bar C^I z\,,
\end{equation}
where there is no implicit sum over $a$ and $b$.
Therefore, if we turn on two fermions with the same $\ell$, off-diagonal contributions arise in the scalar coupling matrix. On the contrary, turning on two fermions with opposite $\ell$ leaves the scalar coupling matrix diagonal.

Since in the main text we restricted our analysis to diagonal scalar matrices, we consider only the case of double fermionic deformations with opposite $\ell$'s, let's say $\ell_a=-\ell_b=\ell$. 

In the present case extra divergent contributions to the vertex come from the field function renormalization. In fact, we recall that in the presence of fermions the scalar coupling matrix renormalizes as
\cite{Castiglioni:2022yes}
\begin{equation}
    (M_I^{\ J})_0= {Z_z}^{-1} \left(M_I^{\ J} + \delta M_I^{\ J} \right) \simeq ( 1 - \delta_z) M_I^{\ J} + \delta M_I^{\ J}  \, ,
\end{equation}
with $\delta_z = -\frac{g^2N}{4\pi\epsilon}\ell(\chi_a^2-\chi_b^2) $. 

Proceeding as done in \ref{sec:pert1}, we impose the matrix counterterm $(\delta M - \delta_z M)$ to cancel divergent contributions
(\ref{eqn:count=0}, \ref{eq:figc1}, \ref{eq:figc2}).  We write
\begin{equation}
\begin{split}
    0=g^2\bigg[ \delta M_I^{\ J} -\frac{g^2 N}{8\pi\epsilon}\big( M_I^{\ K}M_K^{\ J} -\delta_I^J \big)&+ \frac{g^2 N}{4\pi\epsilon}\big( \chi_a^2 + \chi_b^2  \big)\delta_I^J\\
    &-\frac{g^2N}{2\pi\epsilon}\big( \chi_a^2 \delta_I^a\delta_a^J + \chi_b^2\delta_I^b\delta^J_b \big) \bigg]\int d\tau \, \bar z C_J\bar C^I z\,,
\end{split} 
\end{equation}
from which we find
\begin{align}
    (\zeta_a)_0 &=  \left[1+ \frac{g^2N}{4\pi\epsilon}\left( \zeta_a-1 + \ell(\chi_a^2-\chi_b^2) + \frac{(1-\ell)}{2}\frac{(\chi_a^2-\chi_b^2) }{\zeta_a}\right)\right]\zeta_a\,, \\
    (\zeta_b)_0 &= \left[1+ \frac{g^2N}{4\pi\epsilon}\left( \zeta_b-1 + \ell(\chi_a^2-\chi_b^2) - \frac{(1+\ell)}{2}\frac{(\chi_a^2-\chi_b^2) }{\zeta_b}\right)\right]\zeta_b\,, \nonumber \\
    (\zeta_c)_0 &= \left[1 +\frac{g^2N}{4\pi\epsilon}\left(  \zeta_c-1 +\ell(\chi_a^2+\chi_b^2) -\frac{(1+\ell)}{2}\frac{\chi_a^2}{\zeta_c} - \frac{(1-\ell)}{2}\frac{\chi_b^2}{\zeta_c} \right)\right]\zeta_c\,, \quad c\neq a,b \, . \nonumber
\end{align}
Therefore, the $\beta$-functions read
\begin{equation}\label{eq:betasdoubleferm}
\begin{split}
    \beta_{\zeta_a} &= \frac{g^2N}{2\pi}\left( \zeta_a-1 + \ell( \chi_a^2 - \chi_b^2) + \frac{(1-\ell)}{2}\frac{(\chi_a^2-\chi_b^2)}{\zeta_a} \right)\zeta_a\,,\\
    \beta_{\zeta_b}&=\frac{g^2N}{2\pi}\left( \zeta_b-1+\ell( \chi_a^2 - \chi_b^2) -\frac{(1+\ell)}{2}\frac{(\chi_a^2-\chi_b^2)}{\zeta_b} \right)\zeta_b\,,\\
    \beta_{\zeta_c} &= \frac{g^2N}{2\pi}\left(\zeta_c-1 +\ell( \chi_a^2 - \chi_b^2)-\frac{(1+\ell)}{2}\frac{\chi_a^2}{\zeta_c} -\frac{(1-\ell)}{2}\frac{\chi_b^2}{\zeta_c} \right)\zeta_c \, , \quad c\neq a,b \, .
\end{split}
\end{equation}


\section{Enriched flows realization}
\label{app:roadmap}

As highlighted in figure \ref{fig:schematics-enriched}, the three fixed points $W_{1/6}$ and $W_{1/2}^\pm$ can be connected both through enriched flows previously explored in \cite{Castiglioni:2022yes} or through mixed flows as we study here. Naturally, we should be able to recover the former from the latter by specifically tuning our mixed bosonic plus one fermion deformation. In figure \ref{fig:1/2bpsnetwork} we propose a picture that captures the connecting features of each construction.

Concretely, $W_{1/2}^+$ can be obtained from $W_{1/6}$ through a mixed flow in the $(\zeta_2,\chi_1)$ plane and $\ell=1$, as described in section \ref{sec:bosplus1ferm}. An enriched flow, \textit{i.e.} a trajectory that is BPS along all of its points (and not only at the fixed points), is then obtained imposing the constraint $\zeta_2=\chi_1^2$. For generic $\zeta_2$ the corresponding operators are the $1/6$ BPS fermionic loops in green in figure 13 of \cite{Castiglioni:2022yes}, whereas at $\zeta_2=1$ it becomes $W_{1/2}^+$ (see figure \ref{subfig:1/2bpsnetwork-a}).

\begin{figure}[H]
    \centering
    \includegraphics[width=.5\textwidth]{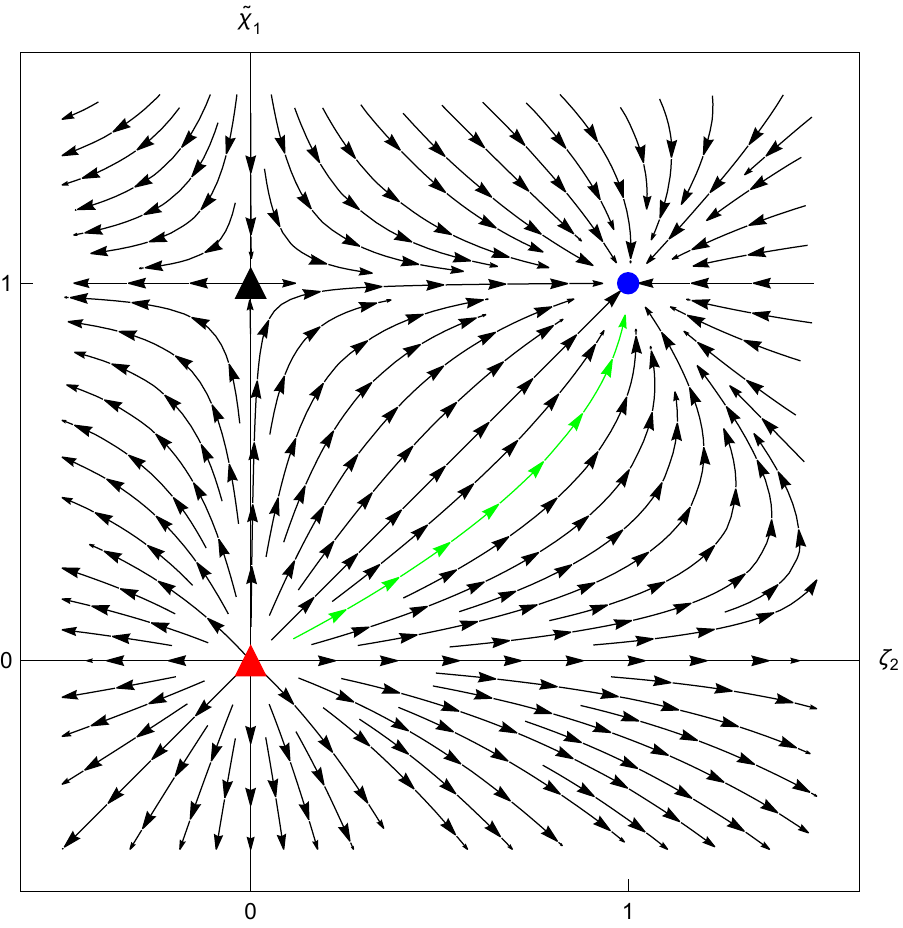}
    \caption{Flows starting from $W_{1/2}^-$. Fixed points correspond to bosonic operators, either $SU(3)$ non-BPS or a $1/6$ BPS $W_{1/6}$. The trajectory in green corresponds to an enriched flow, made by a continuum of $1/6$ BPS fermionic loops.}
    \label{fig:W1/2-toW1/6}
\end{figure}

\begin{figure}[ht]
    \centering
    \subfigure[]{
    \includegraphics[width=0.4\textwidth]{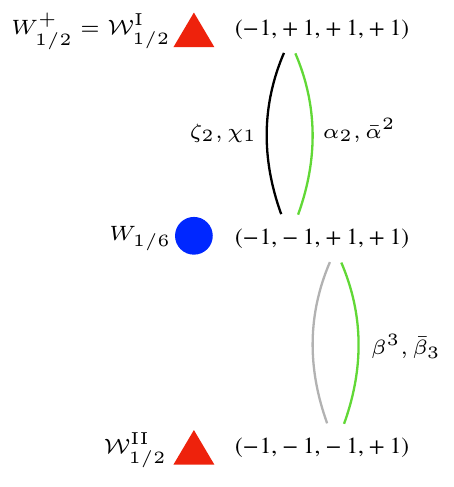}
    \label{subfig:1/2bpsnetwork-a}}
    \subfigure[]{
    \includegraphics[width=0.41\textwidth]{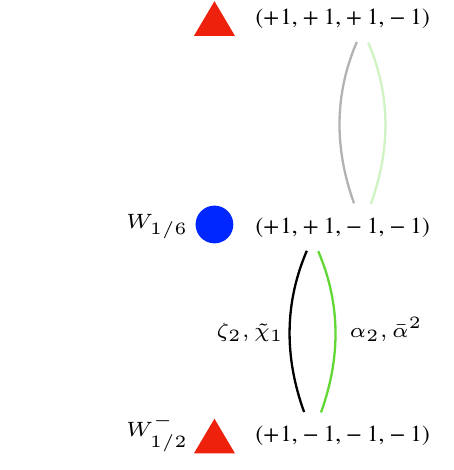}
    \label{subfig:1/2bpsnetwork-b}} 
    \caption{Comparison between enriched RG flows and flows from mixed deformations.}
    \label{fig:1/2bpsnetwork}
\end{figure}

On the other hand, $W_{1/6}$ can be obtained from $W_{1/2}^-$ through a mixed bosonic plus one-fermion flow with $\ell=-1$. We have not considered such a construction explicitly in the main text and for completeness we include it here. The realization involves replicating \eqref{eq:Lfchitilde}-\eqref{eq:betachitilde} with the additional bosonic contribution from $\zeta_2$. Starting with $W_{1/2}^-$ in eq. \eqref{eq:W1/2pm} with $\ell=-1$, we add the mixed deformation
\begin{equation}
    \hat d = -2g^2 \zeta_2 \begin{pmatrix}
        C_2 \bar{C}^2 & 0 \\
        0 & \bar{C}^2 C_2 
    \end{pmatrix} + g \tilde\chi_1 \begin{pmatrix}
        0 & \eta\bar\psi^1 \\
        \psi_1\bar\eta & 0
    \end{pmatrix}\,.
\end{equation}
The corresponding RG flows are then described by the fermionic $\beta$-function in \eqref{eq:betachitilde} together with 
\begin{equation}
\beta_{\zeta_2} = \frac{g^2 N}{2\pi}  \big( (\zeta_2 - 2) - \tilde\chi_1(\tilde\chi_1 - 2) \big) \zeta_2\,.
\end{equation}
We depict the flow in figure \ref{fig:W1/2-toW1/6}. In particular, the enriched flow is obtained imposing the constraint $\zeta_2=-\tilde\chi_1(\tilde\chi_1 -2 )$
and at the particular point where $\zeta_2=1$ we recover $W_{1/6}$ with $M = {\rm diag}(1, 1,-1,-1)$.

Even though by imposing these constraints, either between $\zeta_2$ and $\chi_1$ or between $\zeta_2$ and $\tilde\chi_1$, we do recover enriched flows, these are not precisely the ones studied in \cite{Castiglioni:2022yes}, see figure 1 there. Indeed, $W_{1/2}^+$ is nothing but $\cW_{1/2}^{\rm I}$ introduced there, while $W_{1/2}^-$ coincides with $\cW_{1/2}^{\rm II}$  of \cite{Castiglioni:2022yes} only upon an R-symmetry transformation that interchanges the $1$ and $4$ indices. In fact, if we look at the fermionic couplings, for instance, $W_{1/2}^-$ couples to $\psi_1,\bar\psi^1$ whereas $\cW_{1/2}^{\rm II}$  couples to $\psi_4,\bar\psi^4$. Nevertheless, consistently, under RG flows both $W_{1/2}^-$ and $\cW_{1/2}^{\rm II}$ share the same nature, in the sense that both are repulsive.

The R-symmetry difference between  $W_{1/2}^-$ and $\cW_{1/2}^{\rm II}$ has its origin on a quite technical but nevertheless interesting aspect: from the point of view of enriched flows they are built upon different (but R-equivalent) $1/6$ BPS operators.

To clarify this statement, we recall that the construction for enriched flows is based on the deformation of a $1/6$ BPS bosonic operator. Such a deformation is written in terms of a matrix $G$ that may include all four scalars of the theory via $\alpha$ and $\beta$ parameters,\footnote{This construction was originally proposed in the second chapter of \cite{Drukker:2019bev} and then generalized in \cite{drukker2020bps,Drukker:2020dvr,Drukker:2022ywj,Kong:2022yib}.} 
\begin{equation}
    G = \begin{pmatrix}
    0 && \bar\alpha^1 C_1 + \bar\alpha^2 C_2 + e^{-i \tau} (\beta^3 C_3 + \beta^4 C_4) \\
    \alpha_1 \bar{C}^1 + \alpha_2 \bar{C}^2 + e^{i\tau} (\bar\beta_3 \bar{C}^3 + \bar\beta_4 \bar{C}^4) && 0
    \end{pmatrix}\,.
\end{equation}

In \cite{Castiglioni:2022yes}, the $W_{1/6}$ fixed point around which the deformation is constructed is characterized by a scalar coupling matrix $M=\diag(-1,-1,+1,+1)$ that breaks $SU(4)$ R-symmetry to $SU(2)\times SU(2)$. One can construct BPS deformations which interpolate between $W_{1/6}$ and either $\cW_{1/2}^{\rm I}$ or $\cW_{1/2}^{\rm II}$, according to which $SU(2)$ subgroup the scalars in $G$ belong to. For instance, as shown in figure \ref{subfig:1/2bpsnetwork-a}, if we restrict the parameters in $G$ to be $\alpha_2,\bar\alpha^2$ we obtain a family of 1/6 BPS fermionic WLs interpolating between $W_{1/6}$ and $\cW_{1/2}^{\rm I}$. Instead, if in $G$ we turn on only $\beta^3,\bar\beta_3$ we flow from $\cW_{1/2}^{\rm II}$ to $W_{1/6}$. 
It turns out that  the upper triangle $\cW_{1/2}^{\rm I}$ coincides with $W_{1/2}^+$, instead, as one would naively expect, the lower triangle is not exactly $W_{1/2}^-$.

If we were to find $W_{1/2}^-$ from the enriched flow interpolation, we should consider as one of the two fixed points the 1/6 BPS operator with opposite $M$, as in \cite{Castiglioni:2022yes}. This other flow is depicted in figure \ref{subfig:1/2bpsnetwork-b}. In this case, the BPS interpolation obtained via $\alpha_2,\bar\alpha^2$ would give rise exactly to $W_{1/2}^-$. Note that, in accordance with the map outlined above, from the point of view of mixed deformations, $W_{1/2}^+$ and $W_{1/2}^-$ are both obtained by turning on $\zeta_2$.

Naturally, the flows mentioned above are particular realizations and do not cover all possibilities. In particular, we could also have their $\zeta_3$ counterparts in both \ref{subfig:1/2bpsnetwork-a} and \subref{subfig:1/2bpsnetwork-b}, as well as the $\beta^3,\bar\beta_3$ flow in \ref{subfig:1/2bpsnetwork-b}. We include these flows as light curves to remind that there is such a connection without overloading the picture.

This technical discussion addresses subtleties that should be taken into account when investigating the strong coupling setting. In particular, it follows from the construction of each type of flow that, whereas $\cW_{1/2}^{\rm I,II}$ necessarily share a subset of preserved supercharges, the pair $W_{1/2}^\pm$ does not share any. Thus when studying the corresponding brane/anti-brane description in the dual setting along the lines of \cite{Lietti:2017gtc}, we should keep in mind that mixed flows, and not the enriched ones, are the ones to be considered.

We have discussed one possible scenario of complementary choices of $M$, suitable for turning on enriched flows realized by $\alpha_2,\bar\alpha^2$ (alias $\zeta_2, \chi_1$ mixed flows in this paper) or $\beta^3,\bar\beta_3$ ($\zeta_3, \chi_4$ here). Alternatively, we could consider complementary  $M$ pairs suitable for $\alpha_1,\bar\alpha^1$ or $\beta^4,\bar\beta_4$ ($\zeta_1$ or $\zeta_4$) deformations. In this case we would unravel four extra $1/2$ BPS points that, instead of coupling to fermions with R-symmetry index 1 or 4, would couple to those of index 2 or 3. 

The eight possible realizations of 1/2 BPS fixed points are for instance collected in figure 2(a) of \cite{Lietti:2017gtc}. Referring to that diagrammatic representation, enriched flows interpolate between operators connected by solid red lines (operators that share 2/3 of preserved supercharges), whereas mixed flows connect neighbour operators sharing no common supercharges, denoted there as $W_i$ and $\tilde{W}_i$.

\newpage



\bibliographystyle{JHEP}
\bibliography{refs}

\end{document}